\documentclass[
reprint,
superscriptaddress,
showkeys,
 amsmath,amssymb,
 aps,
pre,
floatfix,
byrevtex
]{revtex4-2}

\usepackage{graphicx}%
\usepackage[english]{babel}
\usepackage{gensymb}
\usepackage{array}[=2016-10-06]
\usepackage{dcolumn}%
\usepackage{bm}%
\usepackage{hyperref}
\hypersetup{
  bookmarksopen = true, 
  bookmarksnumbered = true, 
  colorlinks   = true, %
  urlcolor     = blue, %
  linkcolor    = red, %
  citecolor    = blue %
}

\DeclareUnicodeCharacter{2212}{-}
\usepackage{siunitx}
\DeclareSIUnit\molar{\mole\per\cubic\deci\metre}
\DeclareSIUnit\Molar{\textsc{m}}
\usepackage[version=3]{mhchem} %

\newcommand{\degC}{$\,^{\circ}\mathrm{C}$\,}

\renewcommand{\selectlanguage}[1]{}
\begin{document}

\preprint{APS/123-QED}

\title{This Is Water: Relative Humidity as a Hidden Parameter in the Self-Assembly of CdSe Nanoplatelets into Threads}%

\author{Benoit Wagnon}
\affiliation
{CNRS, ENS de Lyon, LCH, UMR 5182, 69342, Lyon cedex 07, France}

\author{Seungho Lee}
\affiliation
{CNRS, ENS de Lyon, LCH, UMR 5182, 69342, Lyon cedex 07, France}

\author{Lilian Guillemeney}
\affiliation
{ENS de Lyon, CNRS, LCH, UMR 5182, 69342, Lyon cedex 07, France}

\author{Frédéric Chaput}
\affiliation
{CNRS, ENS de Lyon, LCH, UMR 5182, 69342, Lyon cedex 07, France}

\author{Rodolphe Valleix}
\affiliation
{CNRS, ENS de Lyon, LCH, UMR 5182, 69342, Lyon cedex 07, France}

\author{Laurent Coolen}
\affiliation
{Sorbonne Université, CNRS, Institut des NanoSciences de Paris, INSP, F-75005 Paris, France}

\author{Benjamin Abécassis}
\email[To whom correspondence should be addressed: ]{benjamin.abecassis@ens-lyon.fr}
\affiliation
{CNRS, ENS de Lyon, LCH, UMR 5182, 69342, Lyon cedex 07, France}

\date{\today}%

\begin{abstract}
When nanocrystals order into large-scale assemblies, interparticle coupling can give rise to new properties. In the case of semi-conducting nanoplatelets (NPL), stacking them orderly into threads yields highly efficient, ultrafast Förster Resonance Energy Transfer (FRET) over distances of hundreds of nanometers. Producing such assemblies reliably and reproducibly therefore matters. Yet, the mechanism by which dispersed NPL form stacked threads under evaporative self-assembly remains unclear. Here, we show that the relative humidity (RH) under which the NPL dispersion evaporation occurs is a critical, and so far overlooked, experimental parameter. Using a humidity-controlled chamber, we find that micrometer-long NPL stacks form only when evaporation occurs at RH $>$ 50\% rather than under dry conditions and when free oleic acid is present in the dispersion. Because such values lie within typical laboratory conditions, humidity is an easily overlooked parameter. We discuss how adventitious water may modify colloidal forces between NPL, through changes in solvent polarity, depletion, capillary or solvation interactions. Our experiments demonstrate that relative humidity is a key parameter for nanoparticle self-assembly and, more broadly, point to colloidal interactions at the nanoscale. 
\end{abstract}

\maketitle

\begin{figure}[h!]
\centering
\includegraphics[width=0.45\textwidth,keepaspectratio=true]{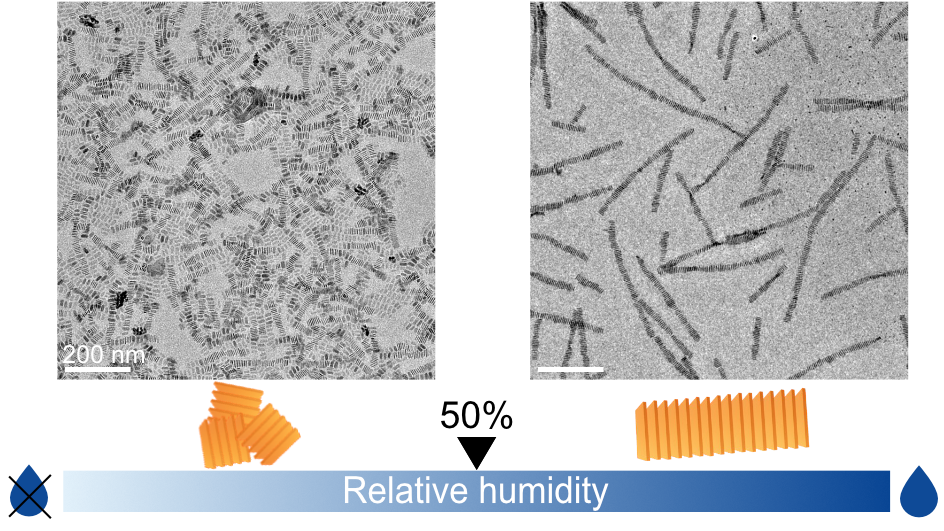}
\end{figure}

\maketitle

\section{Introduction}

Self-assembly is a versatile and efficient process for assembling thousands of building blocks into highly structured objects \cite{Whitesides2002}. In this context, colloidal nanocrystals have established themselves as a model system, combining monodispersity and rich interparticle interactions to yield superlattices whose collective properties resulting from interparticle coupling can surpass that of the individual particles \cite{bassani2024}. Among colloidal nanocrystals, cadmium chalcogenide nanoplatelets (NPLs) \cite{ithurria_quasi_2008, ithurria_colloidal_2011,diroll_2D_2023} stand out owing to their atomically flat morphology and strict thickness quantization, which confer them exceptionally narrow photoluminescence line-widths, giant oscillator strengths, and strongly anisotropic interactions. Various approaches have been used to assemble NPL \cite{guillemeney_Curvature_2022} such as addition of anti-solvent \cite{abecassis_self-assembly_2014}, inclusion in NPL/polymer composites \cite{beaudoin_strain-controlled_2015, beaudoin_reversible_2017}, drying of the dispersion solvent on a solid surface \cite{guzelturk_stacking_2014} or at a liquid/liquid interface \cite{gao_cdse_2017, momper_kinetic_2020, erdem_orientation-controlled_2020}. Threads composed of CdSe NPL stacked face to face form by evaporative self-assembly during which a dispersion of NPL is slowly dried after the addition of oleic acid, yielding highly anisotropic (twisted) threads which can measure up to several micrometers \cite{jana_cdse_2016, jana_ligand-induced_2017}. In this configuration, the NPLs stack face-to-face along their normal axis, separated by a double layer of ligands ($\simeq$ 3.0 nm) attached to the Cd rich basal planes; this close spacing enables ultrafast Förster resonance energy transfer (FRET) between neighbouring platelets \cite{guzelturk_stacking_2014} over distances up to 500 nm \cite{liu_long_2020,ouzit_luminescence_2023,ouzit_fret-mediated_2023}. The threads have transverse dimensions bounded by a single platelet's lateral extent. Depending on how assembly is carried out, the frustration between stacking and particle curvature can be relieved through twisting, yielding chiral threads \cite{jana_ligand-induced_2017,kim_stacking_2019}.  \\

Assemblies results from colloidal interactions \cite{Israelachvili2010} between NPL, a complex interplay between non-additive attractive and repulsive forces acting at the nanometer scale \cite{silvera_batista_nonadditivity_2015, abecassis_three-dimensional_2016} such as van der Waals or dipole-dipole interactions between the cores \cite{dozov2020} or ligand induced forces which often dominate at short ranges compared with the size of the nanoparticles \cite{kister_colloidal_2018, monego_colloidal_2018}, as is the case for NPL. These interactions include a steric short range repulsion which penalizes the interpenetration of ligands from neighbouring nanoparticles, as well as van der Waals attraction between ligands. These forces depend in a subtle fashion on the structure of the monolayer at the atomic scale and can not be properly understood with a mean field view in which sharp interfaces separate the inorganic core, the ligand monolayer and the solvent. Parameters such as the grafting ligand density, their chain length, branching and insaturation, concentration of free ligands in solution impact inter-particle interactions through orientation, ordering and entropic effects \cite{widmer-cooper_orientational_2014, widmer-cooper_ligand-mediated_2016, yang_entropic_2016,kim_surface_2020,winslow_importance_2020}. The solvent in which nanoparticles are dispersed also bears an important role \cite{monego_when_2020}. Solvent penetration inside the ligand monolayer and layering at the monolayer-ligand interface have been shown to be crucial in order to explain the colloidal behavior of NPL beyond the simple solvation picture that considers nanoparticles as atomic solutes \cite{banerjee2026,moretti2026a,Chen2025}. More specifically, solvent layering in the close vicinity of the ligand monolayer induces oscillatory forces whose period and magnitude depend on the chemical structure of the solvent \cite{petersen_crucial_2022,Chen2025}. Whereas solvation forces arise from the ordering of the solvent itself, excluded volume of larger solutes dispersed in it can induce attractive depletion forces \cite{lekkerkerker_colloids_2011,mao_depletion_1995,baranov_assembly_2010}. These interactions can induce nanoparticle assembly when they are mixed with depletants (unbound ligands, micelles, polymer...). Exclusion of these depletants from the inter-particle gap when they approach gives rise to an entropic attractive force. The strength of this interaction depends on the size and concentration of the depletants which would act on the size of the excluded volume and the amount of osmotic pressure acting on the particles respectively. 

The influence of water on nanoparticle self-assembly in organic solvents is an overlooked phenomenon. In apolar solvents such as alkanes, water has very low solubility but their potential enthalpic energy can impact the self-assembly of apolar molecules at the supra-molecular level even at trace level \cite{van_zee_potential_2018}. In the context of quantum dots, water is known to strongly affect their optical properties \cite{hu2021}. Though such effects are known to impact various aspects of (colloidal) science such as lubrication at small scales \cite{feiler2005,boulogne_Cheap_2019}, granular friction \cite{frye2002} or even table tennis racket rubber contact \cite{wu2021}; reports on its impact on nanoparticle assembly in solution are rare \cite{akbulut2007,godfrey_alig_forces_2006,min_frictional_2008,leroch2013,laube2015,chu2025}. When water, even in small quantities, is present in a nanoparticle colloidal solution, capillary bridges can form between particles which induce a strong capillary force and potentially a destabilization of the colloid. This decrease of colloidal stability under humid atmosphere has been noted previously in the case of NPL \cite{jana_stacking_2015}.\\

In the present work, we investigate the formation mechanism of CdSe NPL threads obtained by evaporative self-assembly. Using a custom controlled humidity chamber, we show that the relative humidity of the atmosphere under which NPL colloidal solutions evaporate is a critical experimental parameter. The presence of free oleic acid and a RH above 50\% are both required to afford micrometer-long and single threads. \\

\section{Results and Discussion}

\begin{figure}[htbp]
\centering
\includegraphics[width=0.45\textwidth,keepaspectratio=true]{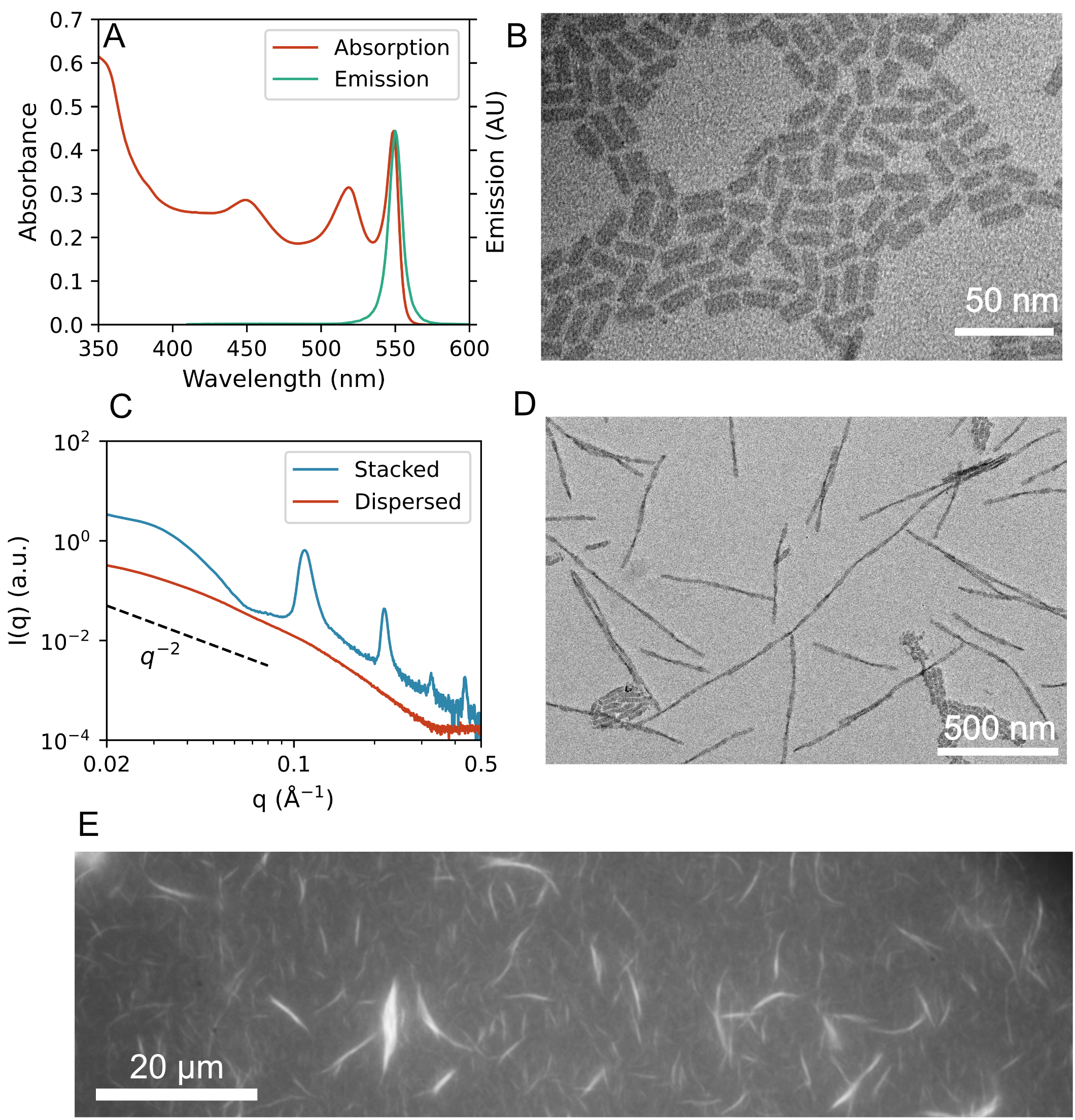}
\caption{\textbf{Characterization of CdSe nanoplatelets and threads:} UV-Vis absorbance and emission spectra (A) and TEM images (B) of the starting 5 ML NPLs. C) SAXS pattern of threads and free NPL. TEM picture (D) and epifluorescence microscopy image (E) of NPL threads.}
\label{NPLs_threads_intro}
\end{figure}

5 monolayers (ML) CdSe NPLs were synthesized, purified, and suspended in hexane following an already established procedure~\cite{ithurria_colloidal_2011, ithurria_quasi_2008} (see Materials and methods section). The NPL thickness and purity are confirmed by absorbance and luminescence spectra (Fig. \ref{NPLs_threads_intro}A) with intense and sharp absorption and emission at $\lambda_{e-hh}$ = 550 nm corresponding to the heavy-hole exciton transition. Bands corresponding to other NPL thicknesses and quantum dots are absent, indicating an efficient NPL isolation. The NPLs are rectangular with an average length of 20.7 nm and a width of 8.3 nm as evidenced by transmission electron microscopy (TEM) (Fig. \ref{NPLs_threads_intro}B). This shape is further confirmed by small-angle X-ray scattering (SAXS) where the characteristic q$^{-2}$ slope is observed ((Fig. \ref{NPLs_threads_intro}C). The NPL surfaces are coated with a mixture of acetate and oleate ligand as evidenced by FT-IR spectroscopy and ${^1}$H NMR (Fig. \ref{threads_IR} and \ref{NPL_NMR}), which endows them with colloidal stability in an apolar solvent~\cite{singh_colloidal_2018-1}. We note the presence of the stretching band of water in the 3400 cm$^{-1}$- 3300 cm$^{-1}$ region and the progression bands associated with the twisting ($\tau$) and wagging ($\omega$) of the methylene units once they are ordered (Table \ref{tab:values_IR_threads} in SI). These bands do not appear in the starting precursor. A low tumbling rate broadens the $^1$H NMR signal, but the broad peak at 5.35 ppm, typical of an alkene, confirms the presence of bound oleate at the NPL surface~\cite{singh_colloidal_2018-1} (Fig. \ref{NPL_NMR}). 

In a typical self-assembly experiment, a small quantity of oleic acid is added to a NPLs dispersion in hexane, which is then slowly dried for 3 days~\cite{jana_cdse_2016, jana_ligand-induced_2017}. (see Materials and methods for more details). Redispersion of the precipitate in hexane yields a turbid colloid, which suggests the presence of large objects. TEM image (Fig. \ref{NPLs_threads_intro}D) reveals thread-like structures that can reach several micrometers in length. Within the threads, the NPLs stack with their widest facets parallel to each other. Since drying artifacts during TEM grid preparation could alter the thread's morphology, we performed cryo-TEM to ensure the reliability of our TEM observations (Figure \ref{Cryo-TEM}). The threads display the same structure when observed with both techniques, but their mean length is shorter in conventional TEM (353 nm versus 538 nm). This shortening likely results from capillary forces during solvent evaporation on the TEM grid. 

Periodicity within the thread is evidenced by a Bragg peak at q$^{*}$ = 0.110 \AA$^{-1}$ in the SAXS pattern which corresponds to a distance (d = $\frac{2\pi}{q^{*}}$) of 5.7 nm between two NPL centers within stacks (figure \ref{NPLs_threads_intro}C) \cite{jana_ligand-induced_2017}. Subtracting the NPL's thickness (1.5 nm for 5 ML NPL) leads to the separation distance between two NPLs of 4.2 nm, slightly higher than two oleic acid molecules in the crystal phase (3.52 nm~\cite{kaneko_structure_1997}). Epi-fluorescence microscopy shows thread-like objects reaching a few micrometers in length (Fig. \ref{NPLs_threads_intro}E), confirming the thread's stability in an apolar solvent and allowing discrimination between stacks and aggregates.

Trials to systematically reproduce this self-assembly protocol yielded seemingly inexplicable variations in thread quality, including length and self-assembly yield. Although all the experimental parameters were kept identical (NPL batch, hexane, and oleic acid sources, glassware, and evaporation times), the outcome varied significantly. Long (µm) well-organized threads were obtained in some instances, whereas the same protocol sometimes yielded precipitates without long-range order. We previously noted variations in colloidal stability of NPL colloids depending on atmospheric humidity: dispersions kept under inert atmospheres were stable while exposure to humid air induced precipitation \cite{jana_stacking_2015}. Relative humidity strongly varies with the weather. For example, yearly meteorological records for a station close to our lab in 2025 shows variations of outdoor RH between 38\% and 100\% with a mean at 73\% $\pm$ 14\% \cite{meteofrance_horaires}. Laboratory RH is further modulated by heating and air conditioning. We hypothesized that these variations could explain the lack of reproducibility of our self-assembly experiments.

\begin{figure*}[htbp]
\centering
\includegraphics[width=0.95\textwidth,keepaspectratio=true]{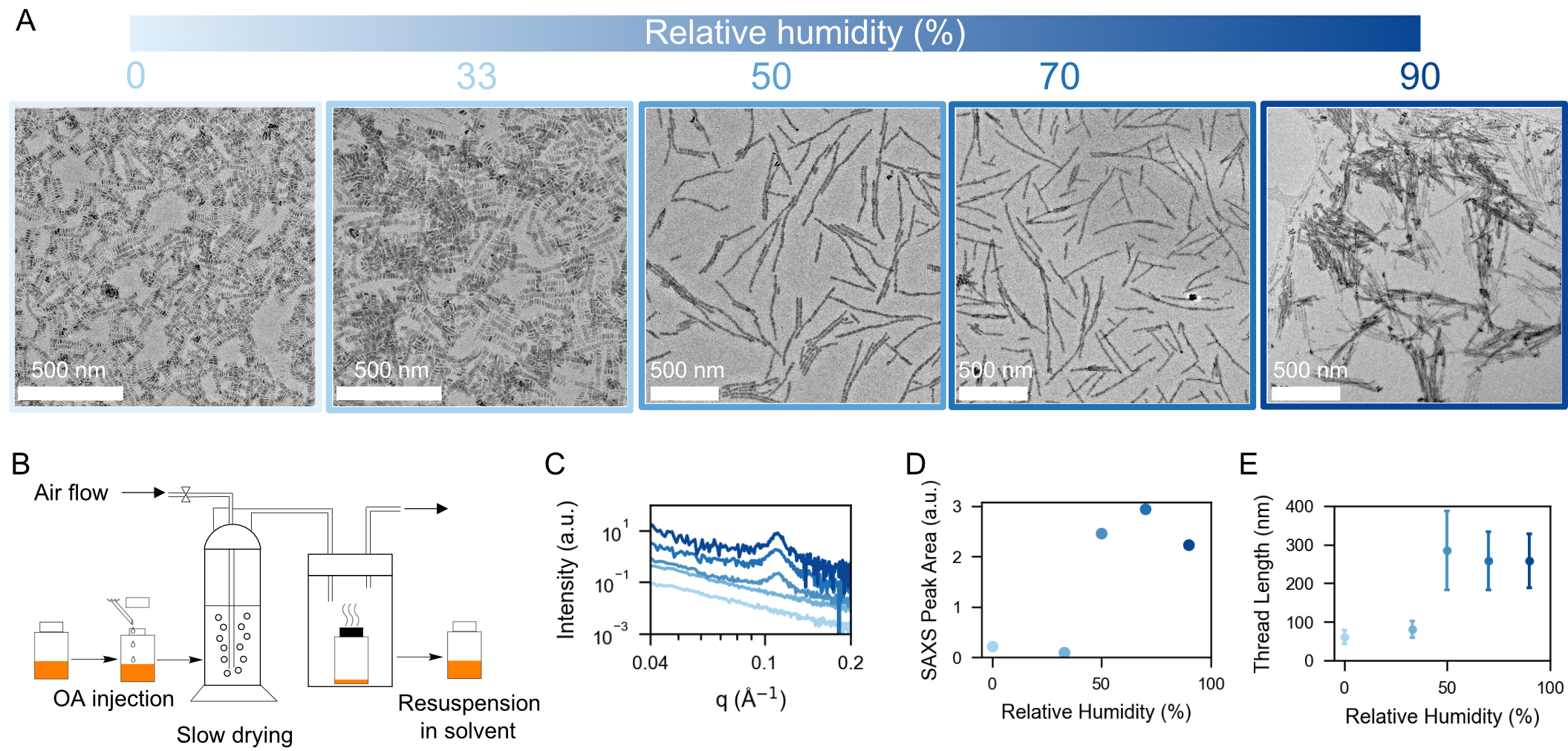}
\caption{\textbf{Effect of relative humidity on CdSe NPL assembly:} A) TEM picture of threads synthesized at different relative humidity (0, 33, 50, 70 and 90\%) at [OA] = 70 mM and three day hexane evaporation. B) Scheme of the experimental setup employed to tune the relative humidity during solvent evaporation. The bubbling solution is composed of water and glycerol at a precise volumetric ratio \cite {forney_control_1992}. C) SAXS pattern of threads suspension synthesized at different relative humidity. D) SAXS peak area and E) mean thread length measured by TEM as a function of the relative humidity (RH) at a [OA] = 70 mM and three day evaporation.}
\label{Humidity_SA}
\end{figure*}

We conducted self-assembly experiments under precisely controlled relative humidity (RH) conditions ranging from 0 to 90\% (Fig. \ref{Humidity_SA}). Measurement at RH = 0\% were performed in a glove box, while experiments at RH between 33\% to 90\% used the setup depicted in Figure \ref{Humidity_SA}B, in which humidity is controlled by bubbling dry air through glycerol-water mixtures of defined volumetric ratios \cite{forney_control_1992}. Figure \ref{Humidity_SA}A reveals a strong influence of RH on the NPL self-assembly. At low RH, TEM images show NPLs predominantly face-down, whereas a sharp transition to an edge-up configuration occurs between 30\% and 50\%. This transition is quantified by the mean thread length $\Bar{L}$ measured in TEM which increases sharply upon crossing this threshold, rising from 80 nm at RH = 33\% to 360 nm at RH = 50\% (Fig. \ref{Humidity_SA}E). Beyond 50\% RH, the mean thread length remains approximately constant, which we attribute to reduced colloidal stability and increased measurement uncertainty arising from side-by-side thread stacking, particularly visible at 90\% RH. The transition between dispersed and stacked NPL observed in TEM is corroborated by SAXS: a stacking peak at q$^{*}$ = 0.110 \AA$^{-1}$ emerges when RH reaches 50\%. Since the NPL concentration is fixed across  experiments, the area under the Bragg peak is proportional to the number of NPL incorporated into the stacks. As shown in figure \ref{Humidity_SA}D, this number increases drastically in humid atmospheres with a sharp transition occurring between 30\% and 50\% RH. Together, both techniques confirm that a high atmospheric water content during solvent evaporation is essential to trigger NPL stacking, and that NPL destabilization is directly correlated with the amount of water vapor present. 

To complement previous reports establishing a link between the mean thread length and the amount of oleic acid (OA) added prior to hexane evaporation\cite{jana_cdse_2016}, we systematically varied the OA concentration at three fixed humidity levels: 33\%, 60\% and 90\% RH. Figure \ref{Humidity_OA_SA}A shows the resulting mean thread length as a function of [OA] and RH. In the absence of OA, no threads were obtained  regardless of humidity, as confirmed by TEM (Fig. \ref{Humidity_OA_SA}A). At 90\% RH, increasing OA concentration produced progressively longer threads,  as previously reported \cite{jana_cdse_2016}, reaching a mean length of 450~nm with individual threads extending up to 2~$\mu$m at 67.6~mM OA (Fig. \ref{Humidity_OA_SA}C). By contrast, at 33\% RH, the threads' mean length plateaus at 175 nm for [OA] above 20 mM. This contrast demonstrates that a water poor environment inhibits thread growth, and that micrometer-long stacks can only be obtained under sufficiently humid conditions irrespective of the OA concentration. Taken together, these experiments prove that both oleic acid and atmospheric water are necessary conditions for the formation of long NPL threads.\\

The formation of long NPL threads requires the concurrent presence of two critical components during evaporative self-assembly: oleic acid and water vapor from the atmosphere. Neither component alone is sufficient, indicating a cooperative mechanism in which both contribute to the thread assembly. Given the very low solubility of water in hexane (around 3.5 mM at 298.31 K~\cite{roddy_solubility_1968, Morgado_solubility_2020}), OA could act as a solubilizing agent that enables water uptake from the atmosphere into the organic phase. Its amphiphilic nature allows it to form reverse micelle-like structures that can encapsulate and transport water molecules within the hydrophobic medium. To verify this hypothesis, we evaporated hexane solutions containing varying amounts of oleic acid under humid atmosphere (90\%RH) and, after 24h, redissolved the residue in the initial volume of hexane before measuring the water content by Karl-Fischer titration (Fig. \ref{Water_uptake_1}). These measurements revealed that the water concentration increases linearly with oleic acid concentration: 0.02 mM of water is solubilized per 1 mM of oleic acid (Fig. \ref{Water_uptake_1}A). The water content reaches saturation after 24 hours of exposure to the humid atmosphere, with no further increase observed at 72 hours (Fig. \ref{Water_uptake_1}B), indicating that our measurements capture the equilibrium water uptake. 

To rationalize our experimental observations, we first discuss the different colloidal forces that water can influence: 1/ Solvent polarity modification: Water uptake into the organic phase increases the medium's polarity, potentially decreasing NPL solubility and thereby promoting aggregation. 2/ Enhanced depletion interactions: Oleic acid could form water-swollen reverse micelles in the organic solvent, creating larger depletant species that induce stronger attractive depletion forces between NPLs. \cite{mao_depletion_1995, lang_introduction_2016, baranov_assembly_2010} 3/ Capillary bridging: Water molecules may accumulate between adjacent NPL surfaces, generating attractive capillary forces through the formation of liquid bridges that pull the particles together. \cite{godfrey_alig_forces_2006, min_frictional_2008} 4/ Surface ligand exchange: Water could facilitate the replacement of initially grafted oleate ligands with hydroxide groups at the NPL surface, diminishing the steric stabilization and reducing colloidal stability.

\begin{figure}[htbp]
\centering
\includegraphics[width=0.45\textwidth,keepaspectratio=true]{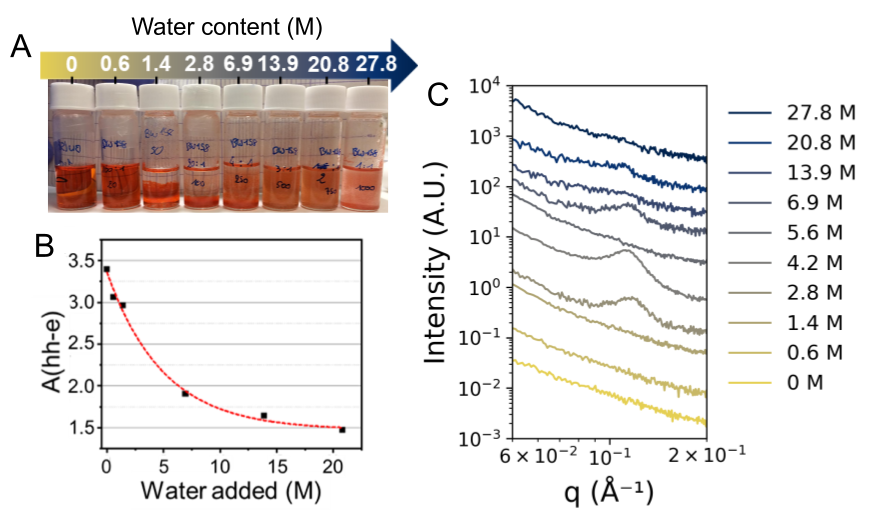}
\caption{\textbf{Effect of water content on NPL dispersions in THF:} A) Picture of vials containing NPLs suspended in THF with increasing amounts of water. B) Evolution of the absorbance at the hh-e transition (A(hh-e)) as more water is added in THF. The red dashed line is an exponential fit of these data. SAXS pattern of the NPL/THF/water suspension for water concentration ranging from C) 0 to 27.78 M.}
\label{NPL_THF}
\end{figure}

We first consider the water-induced solvent polarity change as a cause for NPL stacking. The addition of 4.25 mM water and oleic acid to hexane ($\epsilon \simeq$ 1.88) increases the dielectric constant by less than 0.3\%, representing a very small polarity change. If solvent polarity was the determining factor for NPL stability, one would expect NPLs to be unstable in more polar solvents. However, NPLs dispersed in pure THF ($\epsilon \simeq$ 7.6 i.e. four times higher polarity than hexane) form stable, monophasic suspensions without any aggregation (Figure \ref{NPL_THF}A). The stability persists upon water addition to THF up to water a concentrations of 0.6 M, corresponding to an 11\% polarity increase, approximately three orders of magnitude higher than the water concentration in the thread-forming hexane system (Figure \ref{NPL_THF}A). Above 0.6 M in water, NPLs began to precipitate with characteristic spectroscopic signatures of degradation: exponential decrease in absorbance below 550 nm and appearance of a 570 nm band indicative of quantum dots formed by NPL etching (Figure \ref{NPL_THF}B and Fig. \ref{NPL_THF_UV-Vis}). Increased scattering above 700 nm confirmed the presence of large aggregated objects. The SAXS patterns displayed no stacking peaks up to 2.78 M in water (Figure \ref{NPL_THF}C) but broad stacking peaks, characteristic of defective aggregates rather than ordered assemblies, appear at q* = 0.110-0.119 Å$^{-1}$ (5.28-5.72 nm) for higher water concentrations (2.78-20.83 M). The decreasing SAXS intensity with increasing water content and epifluorescence microscopy observations (Figure \ref{NPL_THF_micro}) both confirm progressive NPL destabilization and aggregation. These results conclusively rule out solvent polarity changes as the driving mechanism for thread formation: NPLs remain stable in high-polarity THF, form threads in low-polarity hexane/OA/water with negligible polarity change, and require water concentrations three orders of magnitude higher in THF to induce disordered aggregation rather than ordered thread assembly.

\begin{figure}[htbp]
\centering
\includegraphics[width=0.45\textwidth,keepaspectratio=true]{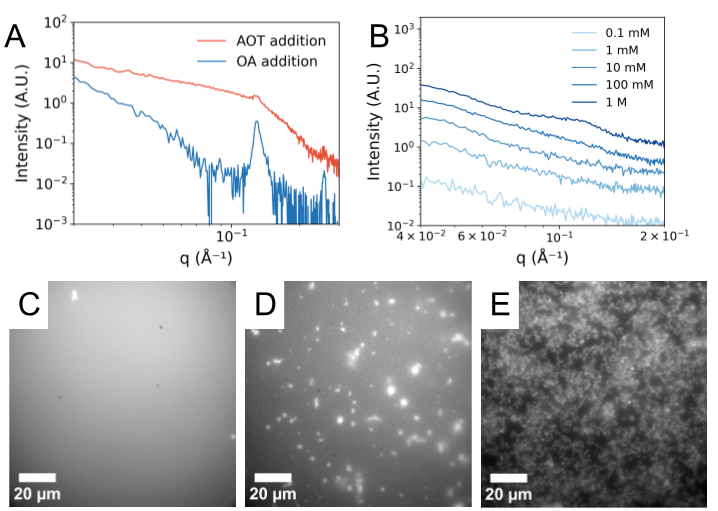}
\caption{\textbf{Comparison between oleic acid, AOT and octadecene in the self-assembly under humid atmosphere:} SAXS diagram of 5 ML NPLs suspension after solvent evaporation at 90\% RH with A) oleic acid (OA) and dioctyl sulfosuccinate sodium salt (AOT) and B) a range of ODE concentration. Epi-fluorescence microscopy pictures of NPLs suspensions with C) AOT, D) OA, and E) 1 M ODE.}
\label{Depletion_interaction}
\end{figure}

Depletion forces are entropic attractions that arise when small solutes (depletants) are excluded from the gap between colloidal particles. The strength of this attraction scales with the depletant concentration and size ~\cite{mao_depletion_1995, lang_introduction_2016,kim2019a}. To test whether free oleic acid molecules or oleic acid reverse micelles act as depletants driving thread formation, we conducted two complementary experiments. First, we replaced oleic acid with octadecene (ODE), a non-amphiphilic molecule of similar molecular weight that cannot bind to NPL surfaces or form micelles. SAXS measurements showed no stacking peaks below 1 M in ODE i.e. two orders of magnitude higher than typical oleic acid concentrations (10 mM) used for thread formation (Figure \ref{Depletion_interaction}B). Even at 1 M, the stacking peak was weak and broad, and epifluorescence microscopy revealed only disordered aggregates rather than threads (Figure \ref{Depletion_interaction}E). Moreover, this mechanism cannot explain water's essential role, as water's concentration is ten times lower than oleic acid and its molecular size is much smaller ($\simeq$0.275 nm vs $\simeq$2 nm). If water-swollen oleic acid reverse micelles were responsible for thread formation through enhanced depletion, then using a surfactant that forms larger micelles should promote stronger assembly. We used dioctyl sulfosuccinate sodium salt (AOT), which forms well-defined water swollen spherical reverse micelles in hexane \cite{Nave2000}. SAXS confirmed spherical AOT micelles with 3.4 nm diameter, larger than any structures observed with oleic acid. Karl Fischer titration showed greater water uptake by AOT after evaporation at 90\% RH (Figure \ref{Water_uptake_AOT}). Despite these larger depletants, SAXS revealed no stacking peaks (Figure~\ref{Depletion_interaction}A), and epifluorescence microscopy showed no threads or aggregation (Figure \ref{Depletion_interaction}C,D), in stark contrast to oleic acid systems. Surface binding of AOT to NPLs is unlikely based on unchanged excitonic absorption, dominant C=O stretching at 1545 cm$^{-1}$ in FT-IR, and AOT's bulky alkyl chain (Figure \ref{AOT_hexane}). These results exclude depletion interactions as the mechanism for thread formation: neither free molecules with similar molecular weight at similar concentrations nor larger reverse micelles induce the ordered assembly observed with oleic acid under humid conditions. 

Solvation forces have been shown to be important for NPL colloidal stability and stacking \cite{petersen_crucial_2022, Chen2025}. In a pure dry solvent, solvation forces depends subtly on the structure of the NPL / solvent interface. They arise form the layering of solvent molecules, with an oscillatory force profiles whose period relates to the bulkiness of the solvent. Interestingly, solvation forces in hydrophobic medium are extremely sensitive to trace amounts of water. Using the surface force apparatus, Christenson \textit{et al.} \cite{christenson1987a,christenson1987} measured a dramatic reduction of the amplitude of oscillatory solvation forces between mica surfaces across octamethylcyclotetrasiloxane when RH increase from 0 to 90\%. They also showed that adhesion at contact, i.e. the force necessary to pull two hydrophobic surfaces once they have been brought in close contact, increases significantly with RH, with a critical value of RH at around 50\% above which the force is two orders of magnitude larger than in the dry case (Figure 3 of \cite{christenson1987a}). This value matches well with our experimental findings with a threshold between ill and well-stacked NPL also at 50\% relative humidity. 

\begin{figure}[htbp]
\centering
\includegraphics[width=0.5\textwidth,keepaspectratio=true]{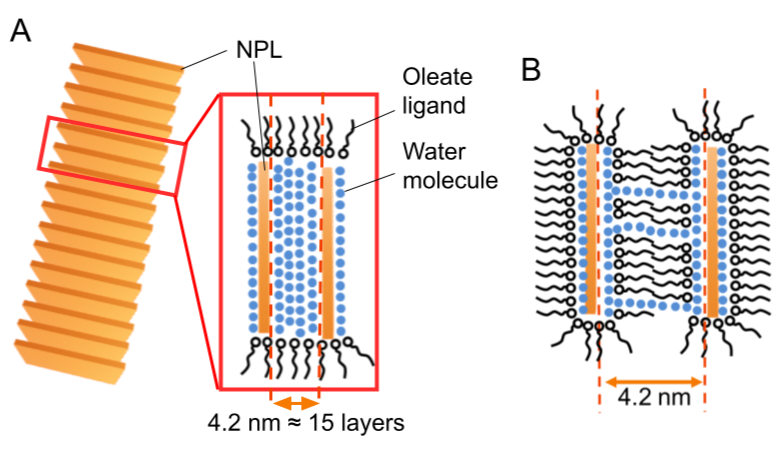}
\caption{\textbf{Schemes of capillary bridging in stacks of NPL in the presence of water:} A) Schematic representation of the water layering in between nanoplatelets in threads structure if formed due to capillary forces. Blue disks correspond to water molecules and black molecules to OA ligands. B) Schematic representation of the water "needles" in between nanoplatelets in threads structure.}
\label{fig:capillary}
\end{figure}

Water brought between the NPL could lead to capillary bridging \cite{godfrey_alig_forces_2006, min_frictional_2008} which would induce a strong attractive force. In this scenario, OA serves as a vessel to transfer water from the atmosphere to the NPL surface vicinity and water fills the gap between NPL (Fig. \ref{fig:capillary} A). A question that arises is thus how water impacts the structure of the ligand monolayer. Changes in the interaction potential between the NPL is likely provoked by this perturbation. We estimate the concentration of water needed to completely fill inter-NPL gaps can be estimated as:
\begin{equation}
C_{\rm H_20}=C_{\rm NPL} \times \frac{A\,d}{v_{\rm H_2O}}, 
\end{equation}
where $A$ is the area of a NPL large facet (20$\times$9 nm$^2$), $d$ is the distance between the NPL surfaces within the stacks (4.2 nm) and $v_{\rm H_2O}$ is the molecular volume of water ($\approx 30 \ \rm{\AA}^3$). This yields $C_{\mathrm{H_2O}}/C_{\mathrm{NPL}} \simeq 2.5 \times 10^{4}$: filling the space between the NPL with pure water would require $\simeq 25 \times 10^{3}$ water molecules per NPL. In our sample, NPL concentration can be estimated to be 2$\times$10$^{-4}$ mM \cite{yeltik_experimental_2015} and the maximum water concentration in OA/hexane solution has been measured to be 5.0 mM so the actual water to NPL ratio is $\simeq 2.4 \times 10^{4}$. The amount of water available is therefore comparable to that required to fill the inter-NPL gaps, making capillary bridging a plausible origin of the attraction. However, that would mean a complete exclusion of the ligand from the gap with water completely replacing it. Furthermore, a single layer of water is about 2.8 \r{A}. Water intercalation would shift the stacking period by around 5\%. This shift is far above the precision of the lamellar period estimation by SAXS which correspond closely to twice the ligands length independent of the experimental conditions. We thus consider that it is unlikely that the ligand is completely replaced by water in the inter-NPL gap. An intermediate scenario in which transient water channel form within the ligand bilayer is also possible. The existence of such (transient) channels is thought to be at the origin of proton conduction in lipid bilayers \cite{marrink1996,lee_dynamic_2024}. However, forming water chains inside hydrophobic alkyl chains bilayers is obviously energetically costly and direct structural observations of needles do not exist. Water might also hydrogen bond to carboxylate oxygen or to Cd surface atoms as a L-type ligand \cite{Owen2015, zherebetskyy_hydroxylation_2014}. Water dissociation at the NPL surface, yielding an hydroxyle (OH$^{-}$) group and a proton which will react with with the COO- group of the native ligand (acetate or oleate), might strongly impact the structure and dynamics of the ligand monolayer and thus change significantly the ligand induced forces between the NPL but also the ligand dynamics at the surface. A significant amount of hydroxyles would show up in the XPS spectra as an additional peak at 531.2 eV in the O 1s region \cite{zherebetskyy_hydroxylation_2014}. XPS experiments performed on pristine NPL and NPL stacks obtained in wet conditions were unable to provide experimental proof of this hypotheses (Figure \ref{XPS_SA}).

To conclude, we investigated the mechanism involved in the self-assembly of colloidal CdSe nanoplatelets and unraveled the critical importance of water and oleic acid for the reproducible assembly of long threads. Using controlled relative humidity setups, we showed that long threads with lengths up to 2 $\mu$m can be consistently obtained when the relative humidity is higher than 50\%. Through a careful screening of the possible mechanisms, we excluded several possible explanations for this surprising phenomenon such as a change in polarity of the solvent or depletion forces induced by water swollen reverse micelles. We offer several perspectives for a comprehensive understanding of atmospheric water influence on evaporative NPL self-assembly such as capillary forces caused by water bridges, or a subtle modification of the solvation forces induced by water induced changes in the monolayer structure. Further studies will aim at understanding such effects in a quantitative fashion. Molecular dynamic simulations would, in this respect, be relevant to unravel the influence of water on atomic arrangements of the NPL/monolayer/solvent interface and their effect on the interaction potential between NPL. We believe such effects are likely to be relevant in nanocrystal self-assembly settings beyond the case of nanoplatelets, as soon as the assembly protocol involves a prolonged contact between non-aqueous dispersions prone to water uptake and the atmosphere.

\begin{acknowledgments}
This article is part of a project that has received funding from the European Research Council (ERC) under the European Union’s Horizon 2020 research and innovation program (Grant agreement No. 865995). This work was supported by the LABEX iMUST of the University of Lyon (ANR-10-LABX-0064), created within the program «Investissements d'Avenir» set up by the french government and managed by the French National Research Agency (ANR) and by the ANR PRC program (project Foenics ANR-20-CE30-0012 and Stefan ANR-25-CE57-1786). The authors thank the CIQLE (Centre d'Imagerie Quantitative Lyon-Est) in Lyon for the availability of the transmission electron microscopy facility. B.A. thanks Asaph Widmer-Cooper for insightful discussions. 
\end{acknowledgments}

\section*{Materials and methods}

\section{Chemicals and instrumentation}
\noindent
\textbf{Reagents}: Sodium oleate (82\%), absolute ethanol, isopropanol (99\%), methanol (99\%), selenium powder (99.5\%), cadmium acetate dihydrate (98\%), methylcyclohexane (99\%), and oleic acid (90\%) were bought from Sigma-Aldrich. 1-octadecene (ODE, 90\%), dioctyl sulfosuccinate sodium salt (AOT) (96\%),  and hexane (90\%) were supplied by Fisher-Agros Organics. Ethyl acetate (99\%) and cadmium nitrate tetrahydrate (98.5\%) were supplied by Fisher-Alfa Aesar. Toluene (99.8\%),THF (99\%) and chloroform (99\%) were bought to Fischer Scientific.\\
Oleic acid was degassed under a vacuum at 120° C for 1 h before being stored on 4 \r{A} molecular sieves under an argon atmosphere to remove water traces. Molecular sieves were also added to hexane for similar reasons. THF was dried using a MBraun SPS5 solvent purification system. All remaining chemicals were used without any further purification.\\

\noindent
\textbf{UV-Vis spectroscopy}
15 $\mu$L of the sample is diluted in 2 mL hexane in a quartz cuvette before any measurement. A JASCO V-730 spectrophotometer was used.\\

\noindent
\textbf{Fourier transform infrared spectroscopy}
FT-IR analyses were carried out on a Perkin-Elmer Spectrum 65 spectrometer in reflection mode on the precursor powders in air.\\

\noindent
\textbf{Fluorescence spectroscopy}
Fluorescence spectra were obtained using a  Horiba-Jobin-Yvon fluorolog-3 fluorimeter using three slits double monochromatic grid system. A continuous  450W xenon lamp generates a non-polarized excitation light. Detection is achieved at a 90° angle by a Hamamatsu R928 (200-900 nm). Samples were diluted by a factor ten and transferred into quartz cuvettes before any measurement.\\

\noindent
\textbf{Epi-fluorescence microscopy}
Epi-fluorescence microscopy pictures were obtained using an Olympus 1$\times$83 equipped with a Hamamatsu ORCA Flash 4 camera, a GFP filter, and a 450 nm blue LED of the pE 300White CoolLED lamp. Pictures were collected and studied by HCImage life and ImageJ, respectively. 0.20$\times$2.00 mm VitroCom flat capillaries were used to enable thread observation in suspension.\\

\noindent
\textbf{Small Angle X-ray Scattering}
A XENOCS Xeuss 3.0 SAXS instrument was used with a X-ray energy of 8 keV. The sample-to-detector distance was set to 350 mm, and a sampling time of 300 s was used to measure scattering. Roughly 100 µL of the NPL solution was inserted in glass capillaries with an overall diameter of 1.5 mm for measurement.\\

\noindent
\textbf{TEM imaging}
TEM pictures were taken with a JEOL JEM-1400 equipped with a 100-kV field emission gun at CIQLE Lyon. One drop of diluted samples was cast on a copper/carbon 300 mesh TEM grid. Nanoplatelets suspension was diluted 10 times to enable a suitable observation.\\

\noindent
\textbf{Karl-fisher measurements}
To estimate the water contents in our solvents after distillation and drying onto molecular sieves, we measured their water content by Karl-Fischer titration using a Mettler-Toledo C20S coulometer coupled to an analytical balance (0.01 mg resolution) and with Hydranal coulomat E as medium. Roughly 0.3 mL of solution was injected three times through the septum using a needled syringe and the mean value was calculated.\\

\noindent
\textbf{Cryo-transmission electron microscopy}
Cryo-TEM imaging was carried out on a JEOL 1400F instrument in Centre Technologique des Micro­structures (CT$\mu$) in University Claude Bernard Lyon 1 (Villeurbanne, France) equipped with a LaB6 electron gun and a precooled cryo-transfer holder (Fischione 2550). 3.5 $\mu$L threads suspension in toluene were deposited on 300 Mesh holey carbon films (C-Flat 2/1) grids and quench-frozen in liquid ethane using a Thermo Fisher Vitrobot. Data was recorded with a Gatan Rio16 camera and a 120-kV accelerating voltage.\\

\noindent
\textbf{Nuclear magnetic resonance (NMR)}
Analyses were performed using a Bruker Advance 300 MHZ spectrometer equipped with a 5mm broadband direct probe BBFO 1H- X Z gradient.

\noindent
\textbf{X-ray photoelectron spectroscopy}
Measurements were made using a THERMO K-alpha+ spectrometer with the Al K$\alpha$ line used as the excitation source at the Laboratoire Science et Surface (SERMA technologies) of Ecully. 

\section{Cadmium precursors, NPL and thread synthesis}
\noindent
\textbf{Cadmium oleate synthesis}
In a 1L three-neck round bottom flask, 12.1776 g (40 mmol) of sodium oleate are dissolved in 200 mL ethanol (96\%) at 70\degC in the presence of 5 mL of water. In parallel, 6.1996 g (20 mmol) of cadmium nitrate tetrahydrate are dissolved in 50 mL ethanol at 96\degC in a conical flask. The cadmium nitrate tetrahydrate solution is then added drop by drop for 30 min. The mixture is left to react for 2 h. Purification is carried out on a Buchner funnel of porosity 4. The solid is washed four times with hot ethanol (96\degC) and two times with methanol. Then, the solid is left to dry under a vacuum overnight before use.\\

\noindent
\textbf{NPL synthesis}
808 mg (1.2 mmol) of cadmium oleate, 54 mg (0.6 mmol) of selenium powder, and 25 mL ODE are inserted in a 50 mL three-neck round bottom flask. This mixture is degassed under a vacuum for one hour. The temperature is then raised to 240°C under argon flow. At 205°C, when the color is yellow-orange, 208 mg (0.8 mmol) of cadmium acetate dihydrate is swiftly injected. The mixture’s color shifted from yellow-orange to deep red. Annealing at 240°C is carried out for 10 min before injecting 1 mL of oleic acid and cooling down the flask to room temperature using a water bath.
NPLs are separated from the remaining reactants and quantum dots by centrifuging the crude mixed with 30 mL hexane and 30 mL isopropanol (in three separate falcons) for 30 min at 6000 rpm. Further purification by centrifugation at 6000 rpm (10 min) is carried out three times in the presence of 4 mL isopropanol to remove the remaining quantum dots. After each centrifugation, NPLs are re-dispersed in 5 mL hexane.\\
An extra purification cycle (centrifugation in the presence of isopropanol and resuspension in hexane) was carried out before FT-IR or NMR measurements.\\

\noindent
\textbf{Thread assembly}
The previously prepared NPLs are mixed with hexane to obtain an absorbance A(hh-e) of 3.6 for a total volume of 2 mL if not stated otherwise. Oleic acid with quantities varying from 6 to 100 µL (11 to 70 mM) is added and a cap with five holes of roughly 1.2 mm in diameter is placed on top before hexane evaporation. The amount of OA is systematically weighed for a more precise measure of its concentration. We used two different set-ups to control the humidity during evaporation. In the first version, the humidity was controlled by bubbling the incoming air in a glycerol-water solution. The relative humidity is then fixed between 10 and 90\% by the water to glycerol ratio \cite{forney_control_1992} and monitored by a hygrometer SUDIPRO. Typical volume fractions used to obtain the different atmospheres are listed in Table \ref{tab:relative humidity} in SI. In a second version (Fig. \ref{fig:RH_regulator_V2}), we used a custom humidity regulator, as described in \cite{boulogne_Cheap_2019}. The humidity regulator operates based on an open-source proportional-integral-derivative (PID) control algorithm programmed on Arduino UNO [https://github.com/br3ttb/Arduino-PID-Library/]. The PID controller equipped with a humidity sensor (Honeywell, HIH-4021-003, 5 V) and a 3-port solenoid valve (SMC V100 series, 12 V) regulates the relative humidity by actuating the valve manifold to mix humid and dry compressed air, thereby maintaining the moisture level in the humidity chamber. Electronic components (TIP 120NPN transistor, 1 and 100 k$\Omega$ resistors, 1N4004 diodes, wires, a 12 V DC power supply) were assembled on a solder-less breadboard according to the specification of the solenoid valve and humidity sensor. Upon entering a user-defined setpoint via the Arduino IDE (https://www.arduino.cc/en/software/), the desired relative humidity was reached and stabilized within 2 hours. After three days of evaporation, the remaining paste is resuspended in 2 mL hexane.\\
Threads were centrifuged three times at 2000 g before any FT-IR, NMR, or XPS measurements.

\bibliography{biblio}

\onecolumngrid
\newpage

\renewcommand{\thefigure}{S\arabic{figure}}
\setcounter{figure}{0}
\setcounter{section}{0}
\setcounter{subsection}{0}

\section{Supplementary materials}

\noindent
\textbf{Influence of water in NPL colloidal stability}\\
70 mM OA was injected in 2 mL of A$_{hh-e}$ = 3.6 5 ML NPL suspension in hexane. Compressed air with a known relative humidity (between 20 and 87\%) was blown gently on the top of the suspension during $\sim$20 s before the vial was tightly closed. 0\% RH was achieved using argon instead of compressed air. Samples were left three days before any measurements.\\

\noindent
\textbf{Determination of water content in hexane/OA solutions}\\
Known OA amounts were added to 2 mL hexane before its evaporation at different times at a 90\% RH. When only OA remained in the vial, 2 mL hexane was added before Karl-Fischer measurement. For water content determination over time, 8 different with identical experimental parameters ([OA]=70 mM, RH=90 \%). Vials were removed from the humidity chamber at different times after the start of hexane evaporation and titrated by Karl-Fischer. \\

\noindent
\textbf{Determination of water content in NPL suspensions}\\
200 $\mu$L of clean NPL batch with A$_{hh-e}$ = 217 was added to 10 mL dried hydranal extracted from the Karl-Fischer apparatus and sealed in a vial. A control vial containing an equal amount of hydranal was prepared. Upon complete NPL dissolution, both solutions were titrated and their difference was calculated to be less than 5 ppm.\\

\noindent
\textbf{Influence of water in NPL/THF suspensions}\\
In 5 ML NPL in THF suspension, 0 to 1 mL water was added to yield a 2 mL A$_{hh-e}$ = 3.6 suspension. The cap was tightly closed and samples were left for 3 days before measurements.\\

\noindent
\textbf{ODE to trigger NPL self-assembly}\\
An increasing amount ( 0.1 to 1000 mM) of 1-octadecene (ODE) was added to NPL in hexane to form 2 mL of A$_{hh-e}$ = 3.6 suspension before 3 days evaporation at 90\% RH. The remaining paste was resuspended to obtain a 2 mL suspension before measurements\\

\noindent
\textbf{Determination of NPL surface chemistry when in contact with AOT}\\
71 mM AOT amount was added to 2 mL of A$_{hh-e}$ = 3.6 5 ML NPL suspension in hexane. After 3 days, UV-Vis and emission spectra were recorded. The suspension is then centrifuged three times in the presence of ethanol to purify the particles before FT-IR measurement.\\

\noindent
\textbf{Determination of water content in hexane/AOT solutions}\\
71 mM AOT amount was added to 2 mL hexane before its evaporation at different times at a 90\% RH. When only OAT remained in the vial, 2 mL hexane was added before Karl-Fischer measurement.\\

\noindent
\textbf{Comparison between OA and AOT to trigger NPL self-assembly}\\
70 mM of surfactant was injected in 2 mL of A$_{hh-e}$ = 3.6 5 ML NPL suspension in hexane before 3 days evaporation at 90\% RH. The remaining paste was resuspended in 2 mL hexane before measurements.\\

\noindent
\textbf{SAXS of AOT reverse micelles}\\
70 mM AOT and OA solution in hexane have been measured in SAXS glass capillaries to determine their micelle structure.\\

\section{Supplementary figures and tables}

\begin{figure}[htbp]
\centering
\includegraphics[width=1\textwidth,keepaspectratio=true]{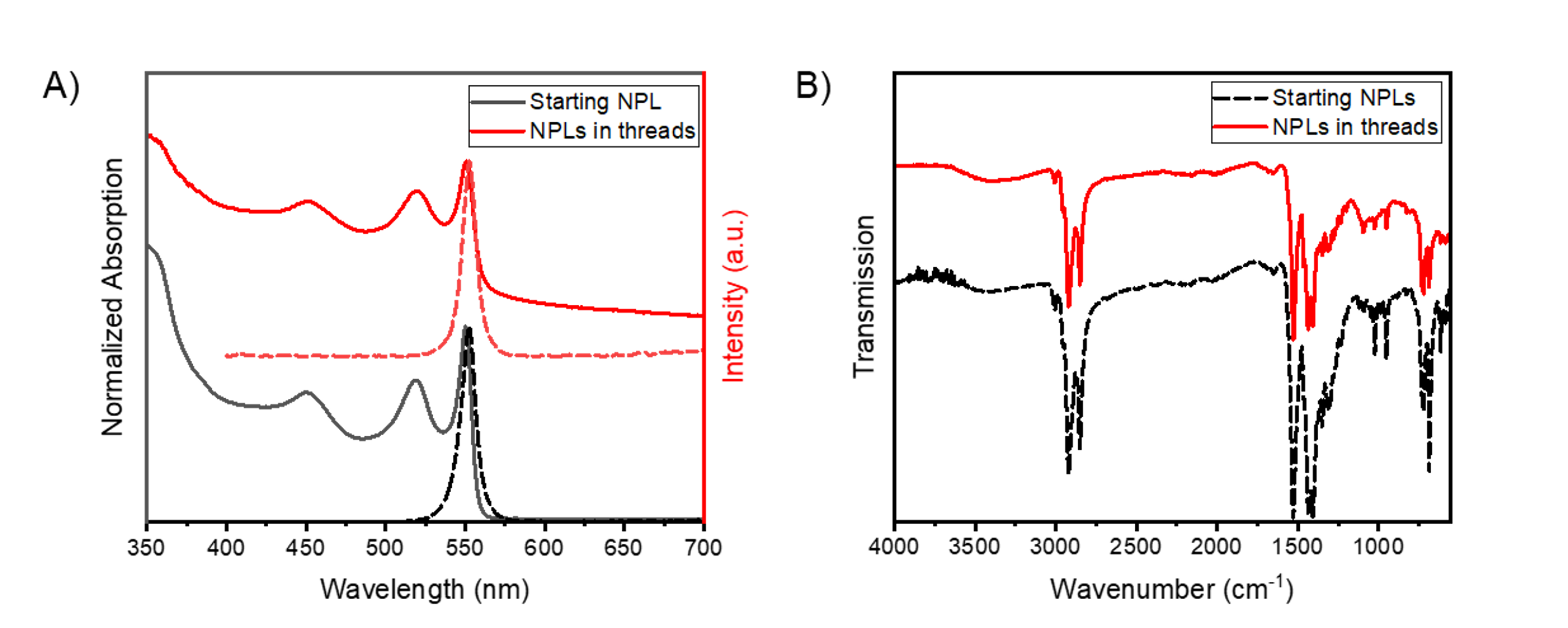}
\caption{A) UV-Vis and emission spectra of the starting NPLs (in black) and threads (in red). B) The corresponding FT-IR spectra (same color code used). For both graphs, the curves are superposed for clarity reasons.}
\label{threads_IR}
\end{figure}

\begin{figure}[htbp]
\centering
\includegraphics[width=0.8\textwidth,keepaspectratio=true]{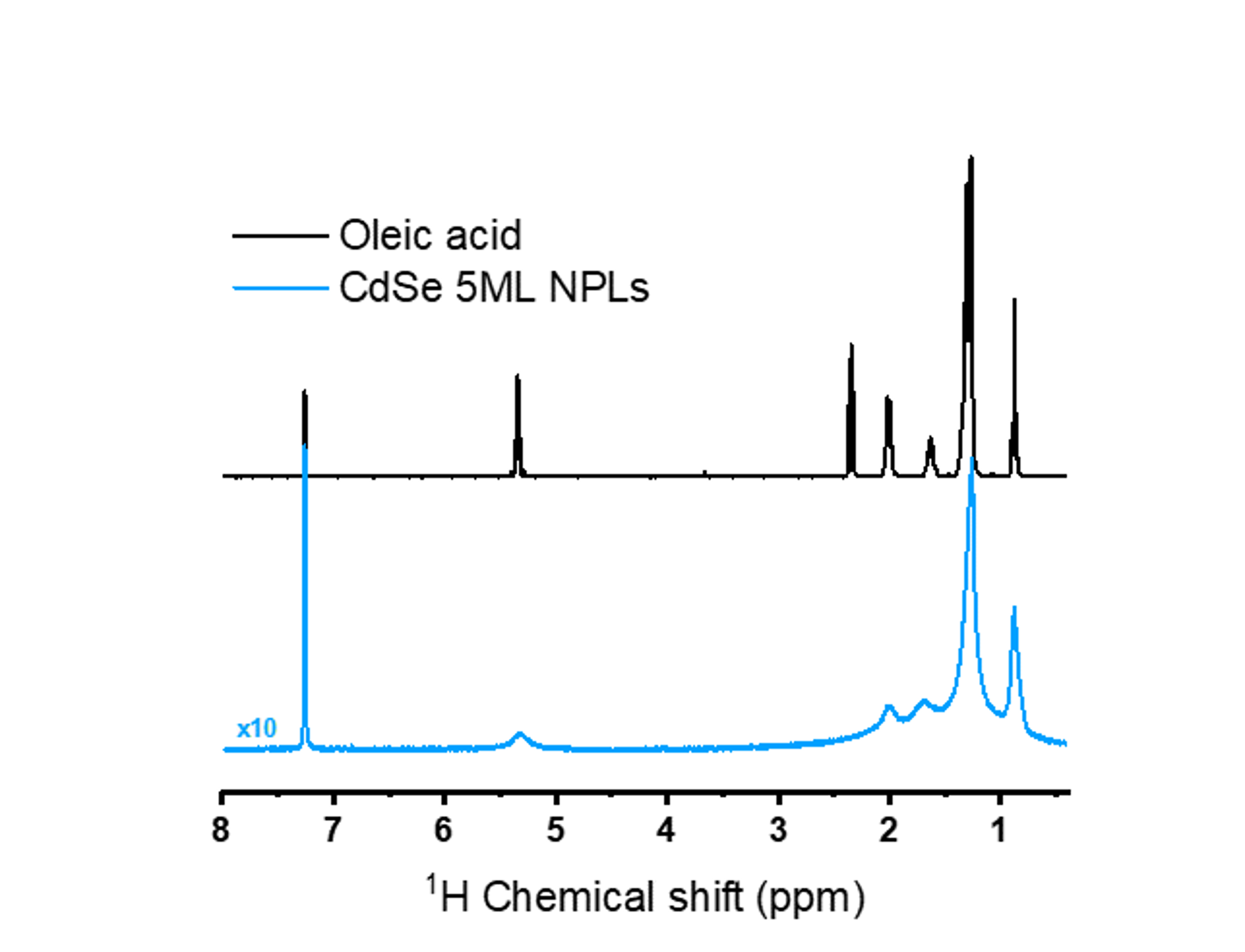}
\caption{NMR spectra of oleic acid (in black) and NPL coated with the oleic acid/acetate mixture (in blue). Spectra are shifted for clarity and the NPL spectrum have been magnified ten times.}
\label{NPL_NMR}
\end{figure}

\begin{figure}[htbp]
\centering
\includegraphics[width=0.8\textwidth,keepaspectratio=true]{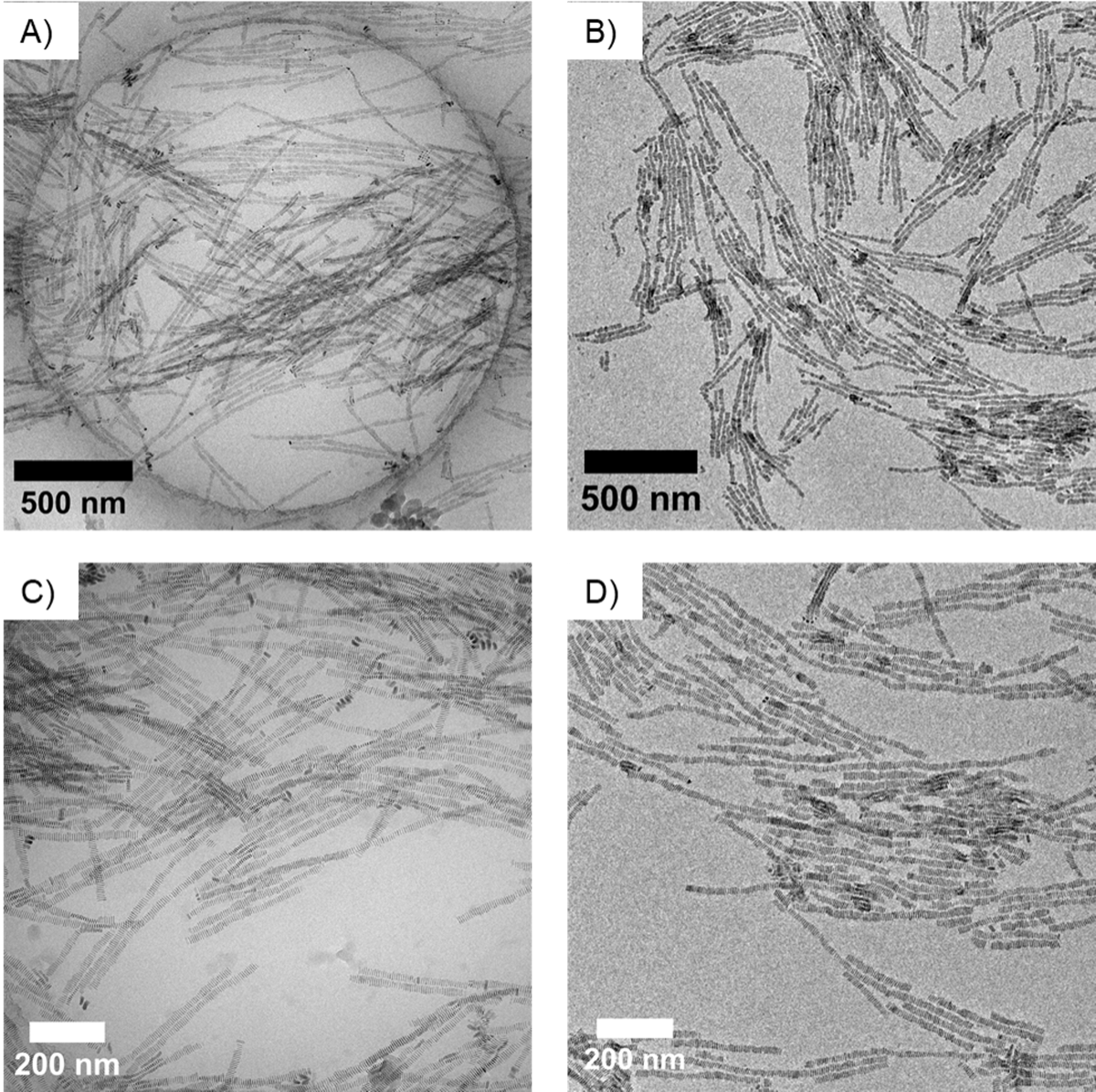}
\caption{Cryo-TEM (left column) and TEM (right column) picture of 5 ML NPLs threads. A), B) and C), D) are comparisons at a low and high magnification respectively. Thread morphology is independent of the observation method when features such as NPLs bend, spacing, and thread defects are considered. There are two notable differences: 1) the absence of NPLs lying on their widest face in the cryo-TEM picture and 2) a slight decrease in the mean thread length from 538 $\pm$ 292 nm to 353 $\pm$ 225 nm for cryo-TEM and TEM observation respectively.}
\label{Cryo-TEM}
\end{figure}

\begin{figure}[htbp]
\centering
\includegraphics[width=0.8\textwidth,keepaspectratio=true]{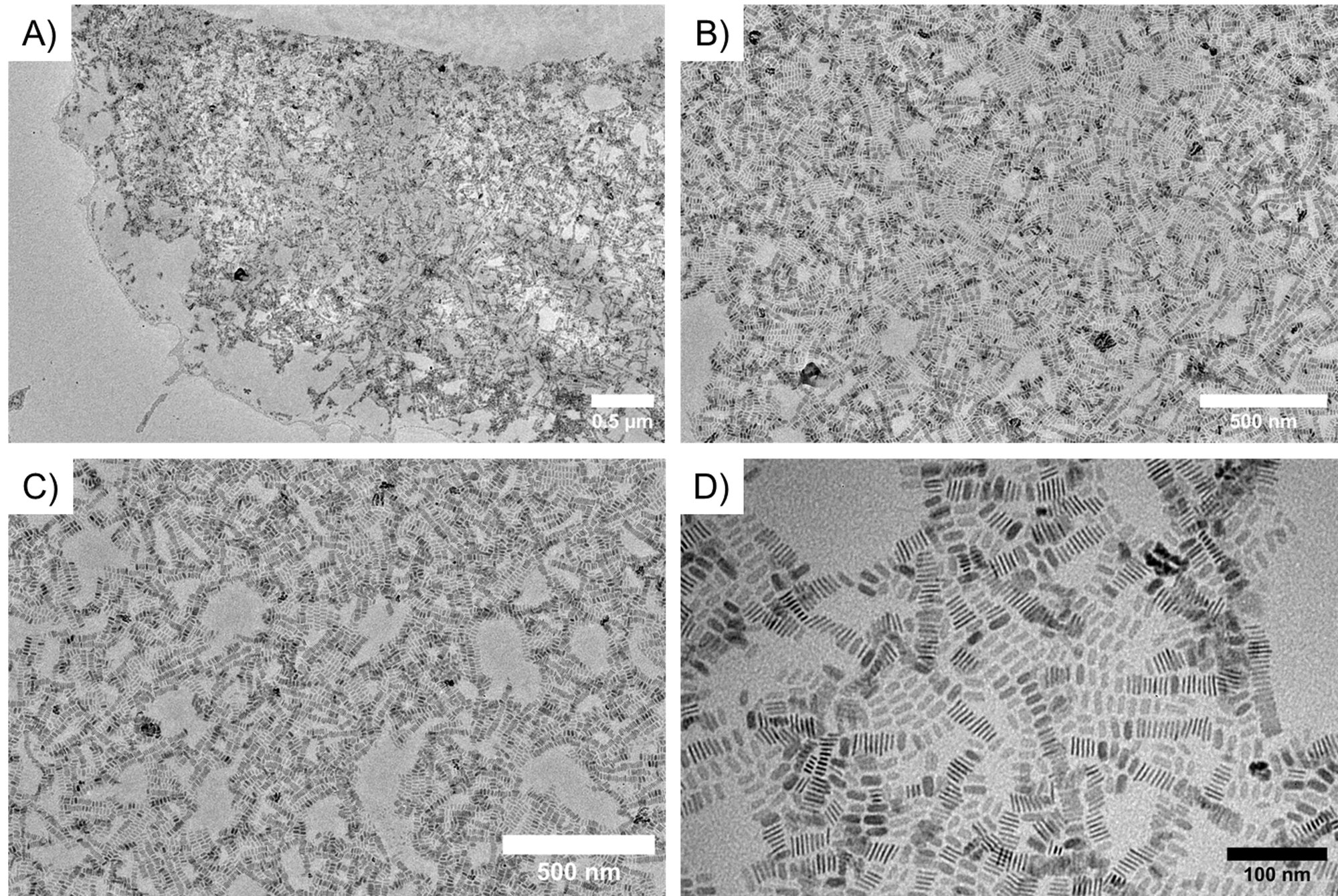}
\caption{TEM pictures at several magnification of threads sample at 0\% RH with [OA] = 70 mM injected before a three day hexane evaporation.}
\label{TEM_RH0}
\end{figure}

\begin{center}
\begin{table}[htbp]
\caption{Frequency values in the 1600-1400 cm$^{-1}$ and the CH$_2$ vibration mode for OA-capped free NPLs and NPL in threads. All units are in cm$^{-1}$.}
\begin{tabular}{cccccc}
\hline
Compounds & $\nu_{as}$(COO$^{-}$)  & $\nu_s$(COO$^{-}$)   & $\Delta\nu$(COO$^-$) & $\nu_{as}$(CH$_2$)  & $\nu_s$(CH$_2$) \\
\hline
Free NPL & 1530 & 1406 & 124 & 2921 & 2851\\
NPL in threads & 1527 & 1407 & 120 & 2920 & 2850\\
\hline
\end{tabular}
\label{tab:values_IR_threads}
\end{table}
\end{center} 

\begin{table}[h]
    \centering
    \begin{tabular}{c|c|c}
        Relative humidity (\%) & Glycerol volume (mL) & Water volume (mL) \\
         90 & 0 & 100 \\
         70 & 53 & 47 \\
         60 & 66 & 41 \\
         25 & 90 & 10
    \end{tabular}
    \caption{Typical glycerol and water volumes mixed to obtain desired relative humidity}
    \label{tab:relative humidity}
\end{table}

\begin{figure}[htbp]
\centering
\includegraphics[width=0.8\textwidth,keepaspectratio=true]{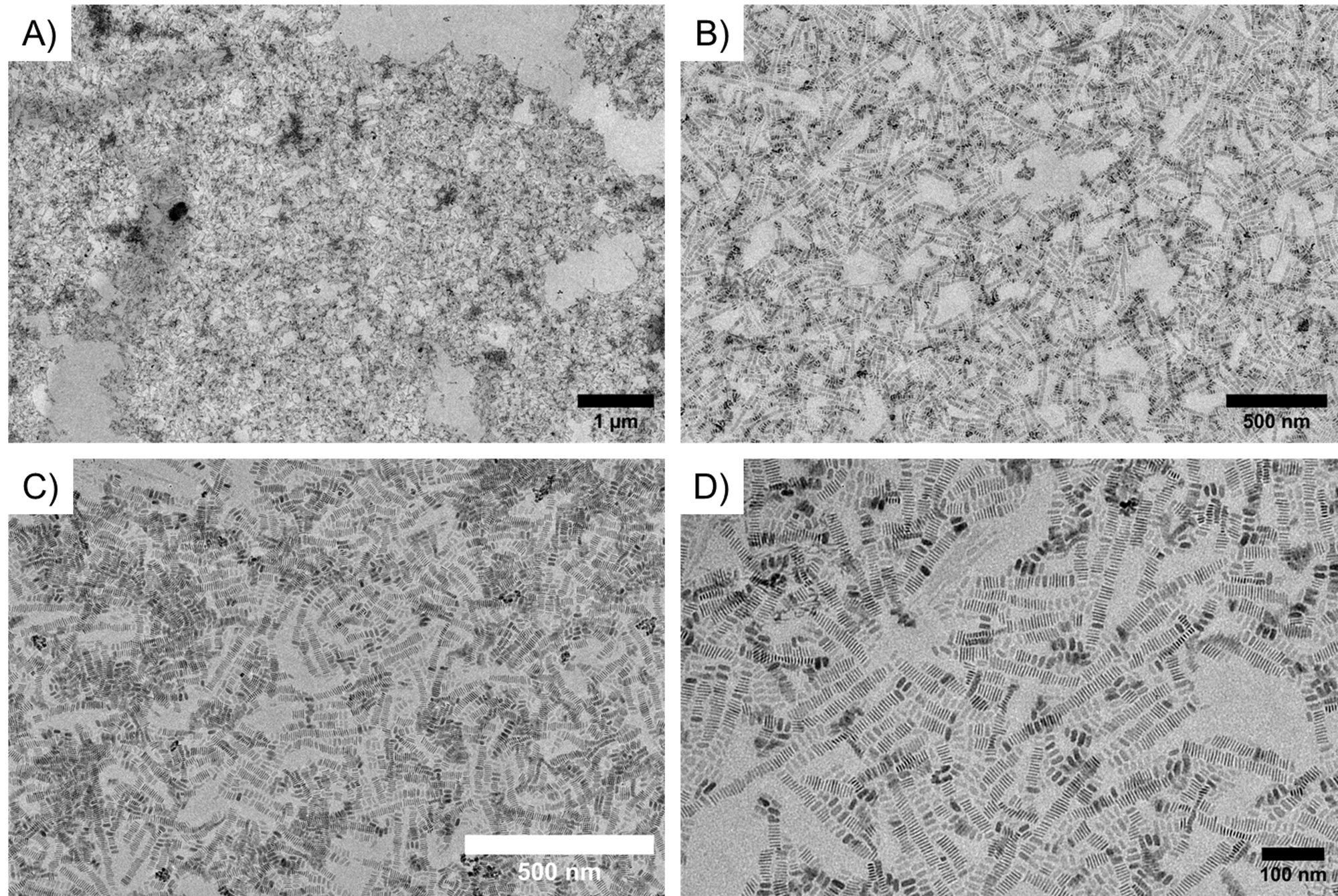}
\caption{TEM pictures at several magnification of threads sample at 33\% RH with [OA] = 70 mM injected before a three day hexane evaporation.}
\label{TEM_RH33}
\end{figure}

\begin{figure}[htbp]
\centering
\includegraphics[width=0.8\textwidth,keepaspectratio=true]{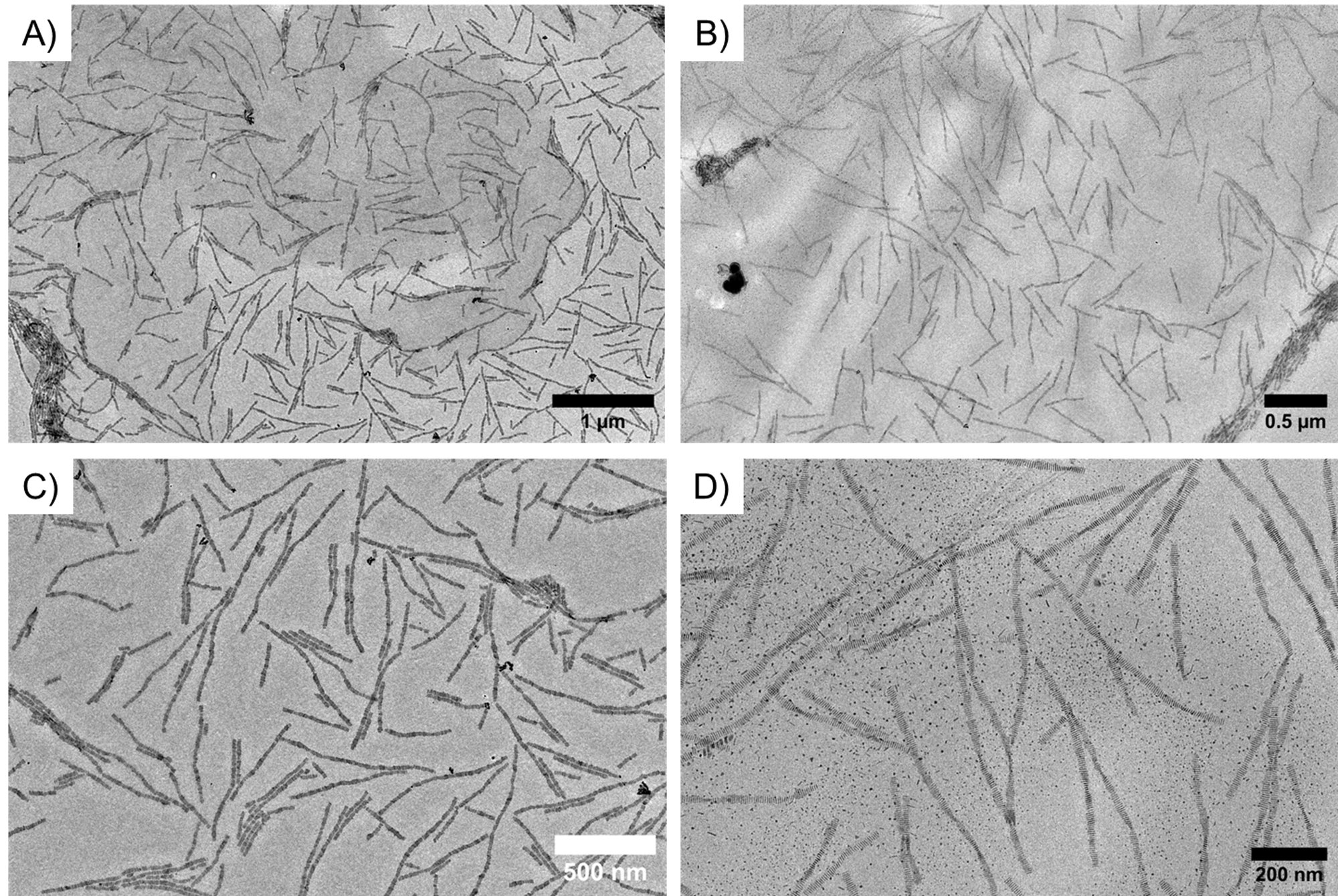}
\caption{TEM pictures at several magnification of threads sample at 50\% RH with [OA] = 70 mM injected before a three day hexane evaporation.}
\label{TEM_RH50}
\end{figure}

\begin{figure}[htbp]
\centering
\includegraphics[width=0.8\textwidth,keepaspectratio=true]{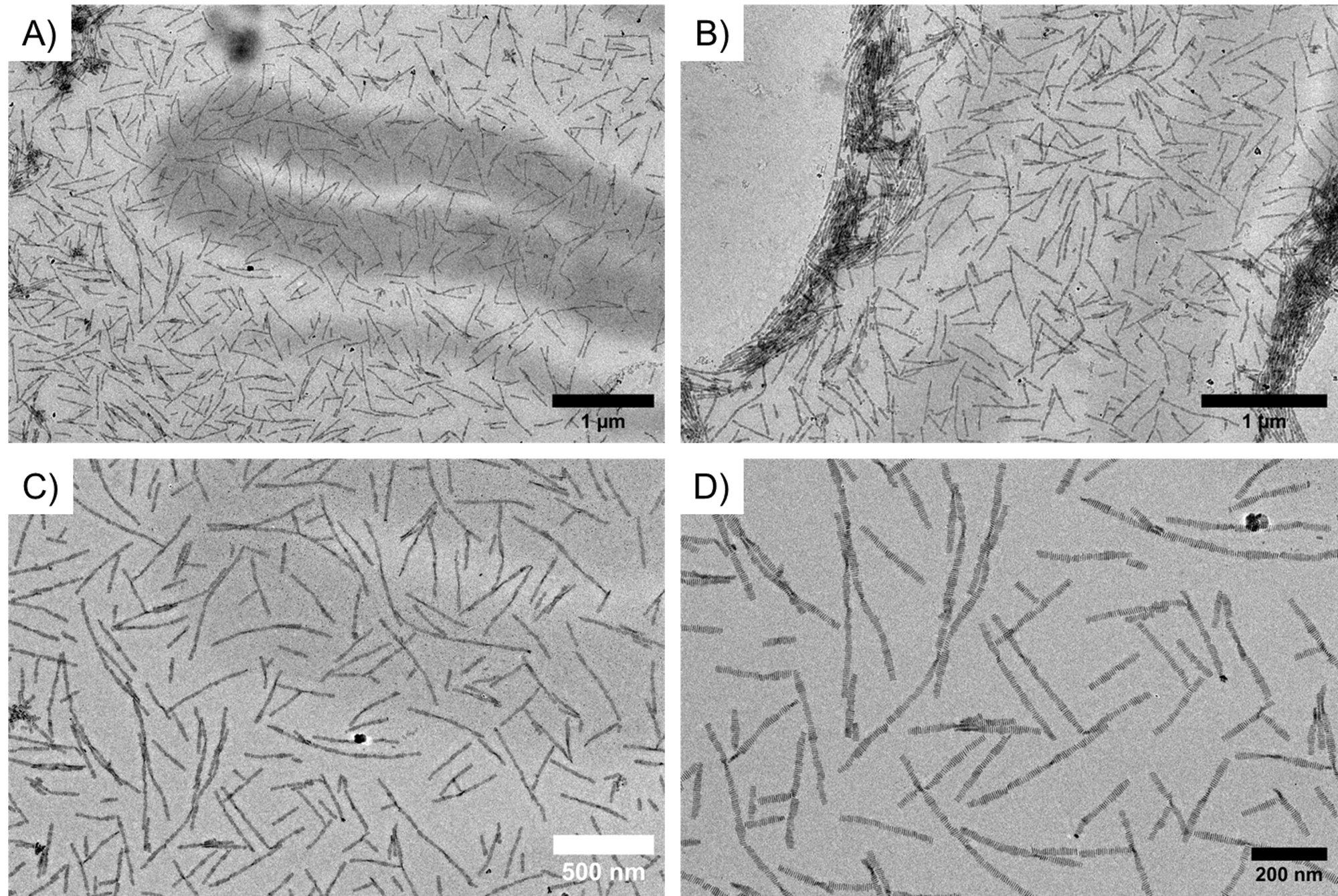}
\caption{TEM pictures at several magnification of threads sample at 70\% RH with [OA] = 70 mM injected before a three day hexane evaporation.}
\label{TEM_RH70}
\end{figure}

\begin{figure}[htbp]
\centering
\includegraphics[width=0.8\textwidth,keepaspectratio=true]{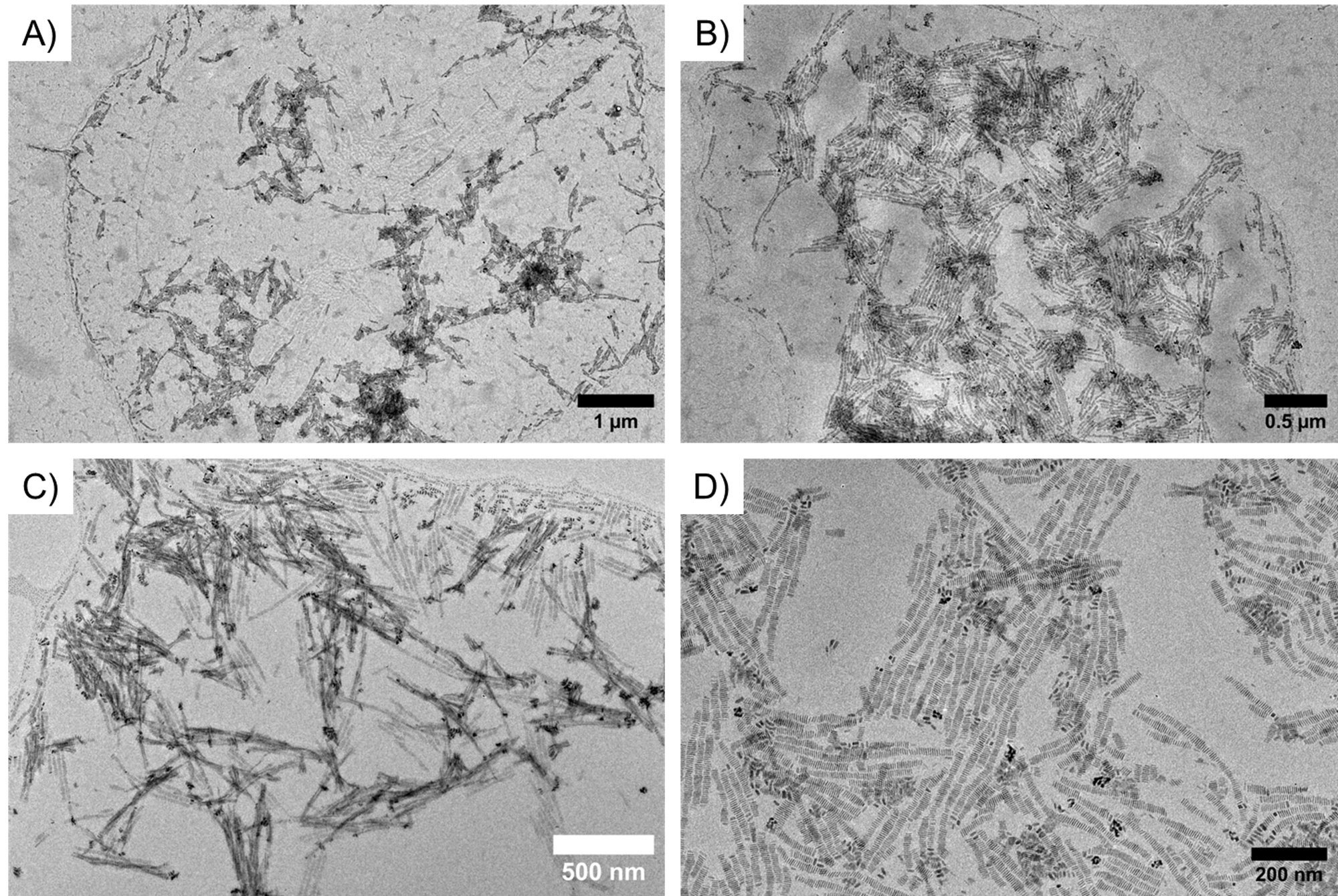}
\caption{TEM pictures at several magnification of threads sample at 90\% RH with [OA] = 70 mM injected before a three day hexane evaporation.}
\label{TEM_RH90}
\end{figure}

\begin{figure}[htbp]
\centering
\includegraphics[width=0.8\textwidth,keepaspectratio=true]{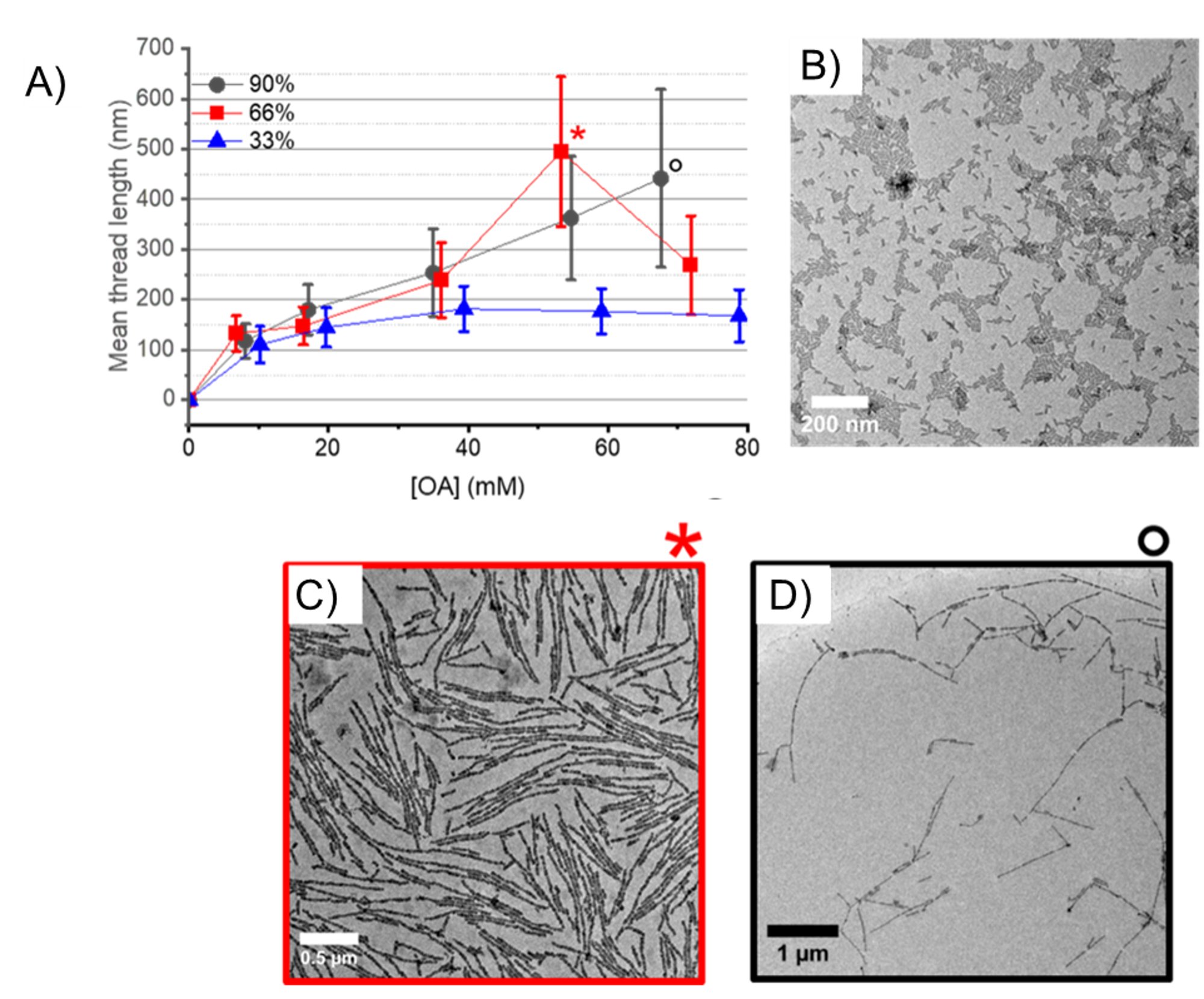}
\caption{A) Mean thread length measured by TEM as a function of the injected OA concentration for three different relative humidity. TEM pictures of samples with B) no OA injected C) corresponding to 55 mM OA injected and evaporation at 66\% RH and D) 70 mM OA injected and hexane evaporation at 90\% RH.}
\label{Humidity_OA_SA}
\end{figure}

\begin{figure}[htbp]
\centering
\includegraphics[width=1\textwidth,keepaspectratio=true]{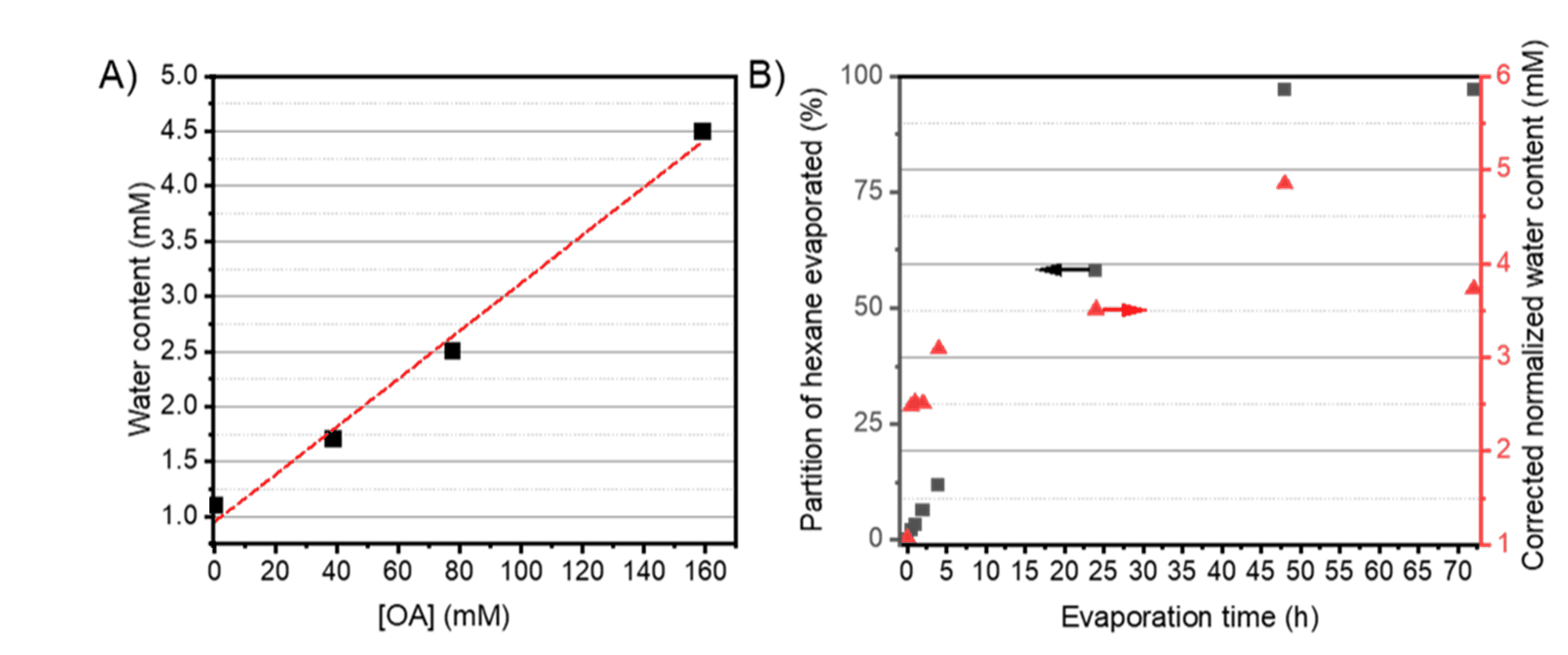}
\caption{A) Water content in a resuspended hexane/OA with increasing OA concentration, after OA injection and 24h of hexane evaporation, similar to the thread synthesis process. The red dashed line is a linear fit. B) Mass loss (black square) and water uptake (red triangles) by hexane/OA solutions during solvent evaporation. OA concentration is fixed at 70.5 mM for all 8 samples and the water content is corrected depending on the mass lost during the evaporation process and the mass of OA injected compared to the sample that didn't undergo solvent evaporation.}
\label{Water_uptake_1}
\end{figure}

\begin{figure}[htbp]
\centering
\includegraphics[width=0.8\textwidth,keepaspectratio=true]{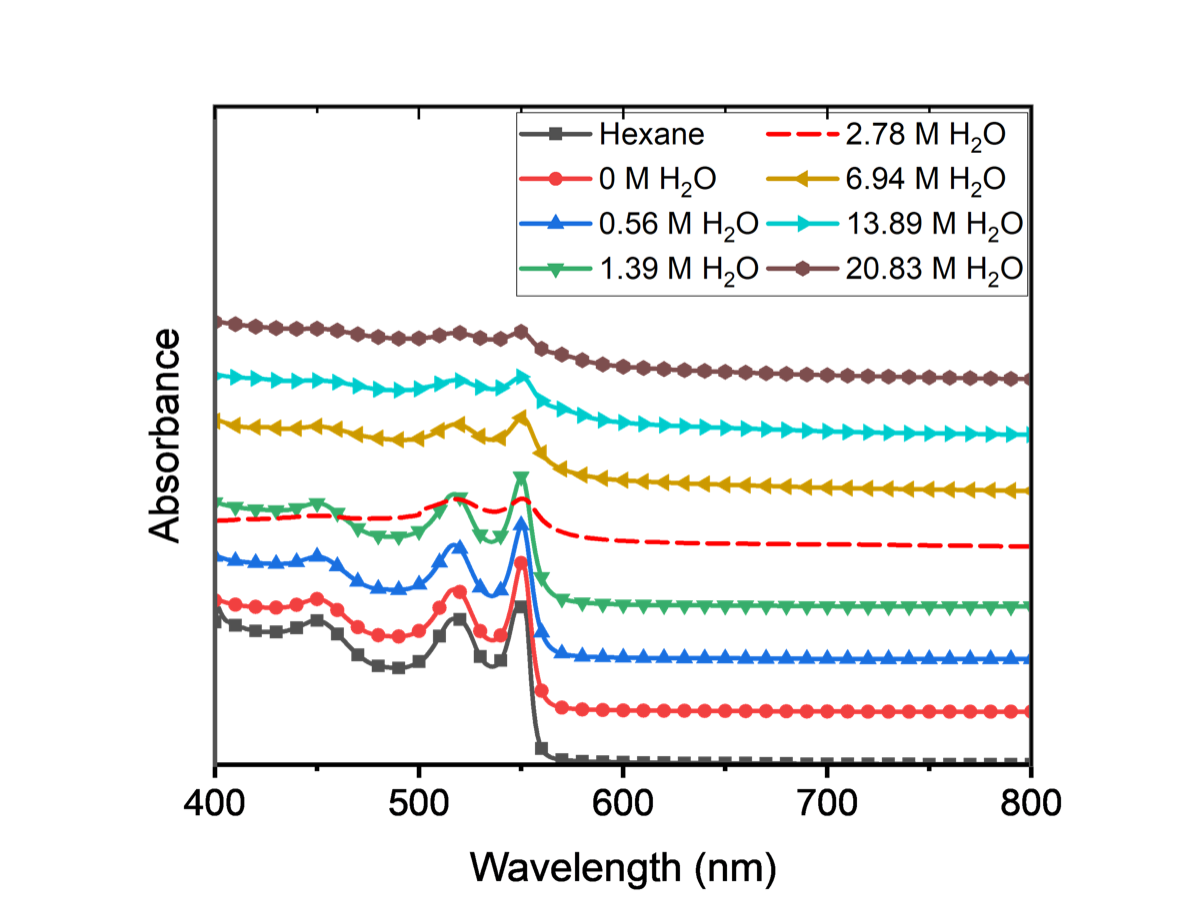}
\caption{UV-Vis spectra of the NPL/THF/water suspension in figure \ref{NPL_THF} along with a reference in hexane. We can see a decrease in the hh-e transition absorption and increased scattering. This indicates both NPL degradation and aggregation take place. The only exception is the sample containing 2.78 M water (red dashed line) that flocculated before the UV-Vis spectra measurement was finished. Curves are superposed for clarity.}
\label{NPL_THF_UV-Vis}
\end{figure}

\begin{figure}[htbp]
\centering
\includegraphics[width=0.9\textwidth,keepaspectratio=true]{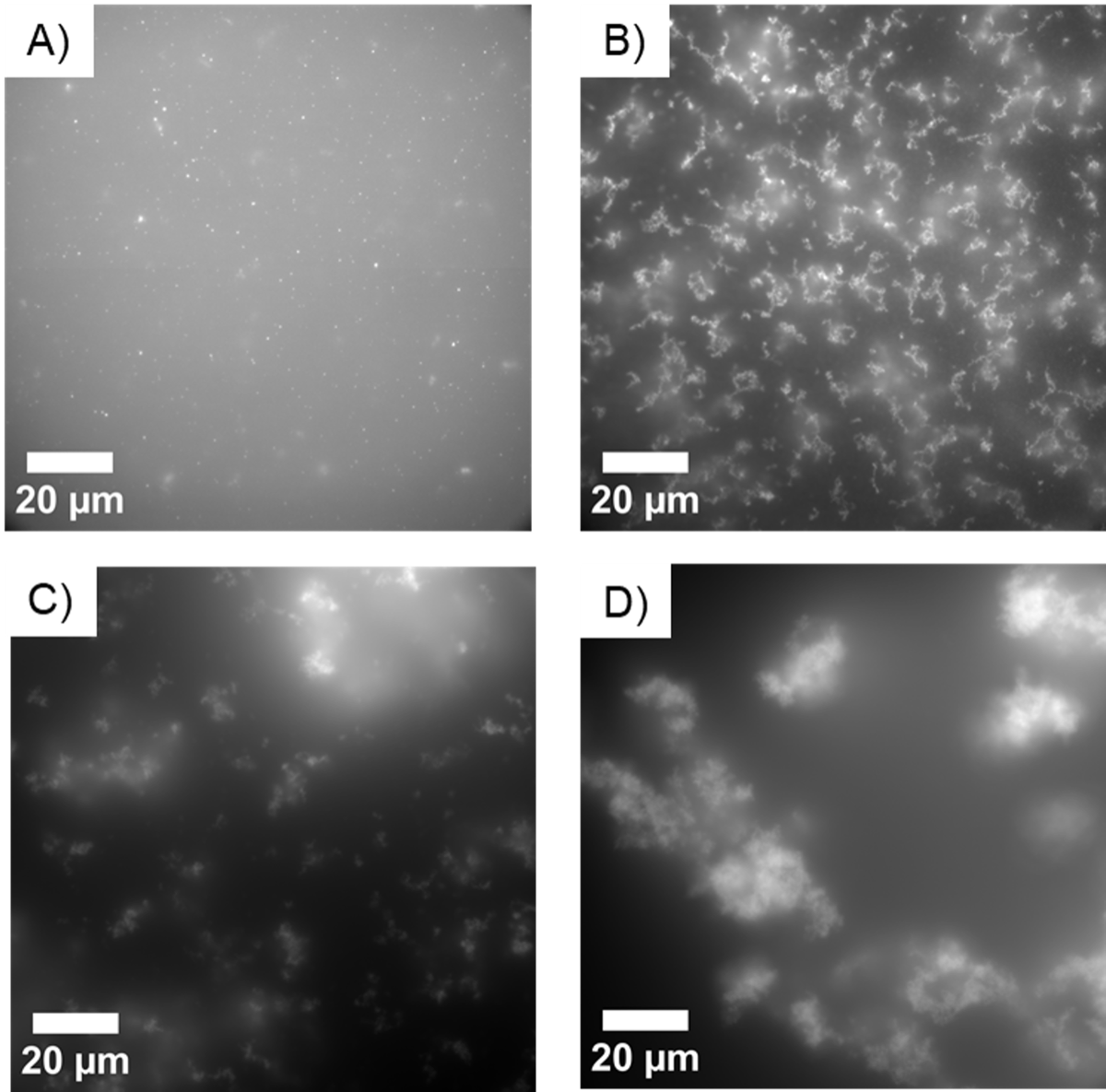}
\caption{Epi-fluorescence microscopy picture of NPL/THF suspensions with A) 0 M, B) 1.39 M, C) 2.78 M, and D) 20.83 M water.}
\label{NPL_THF_micro}
\end{figure}

\begin{figure}[htbp]
\centering
\includegraphics[width=1\textwidth,keepaspectratio=true]{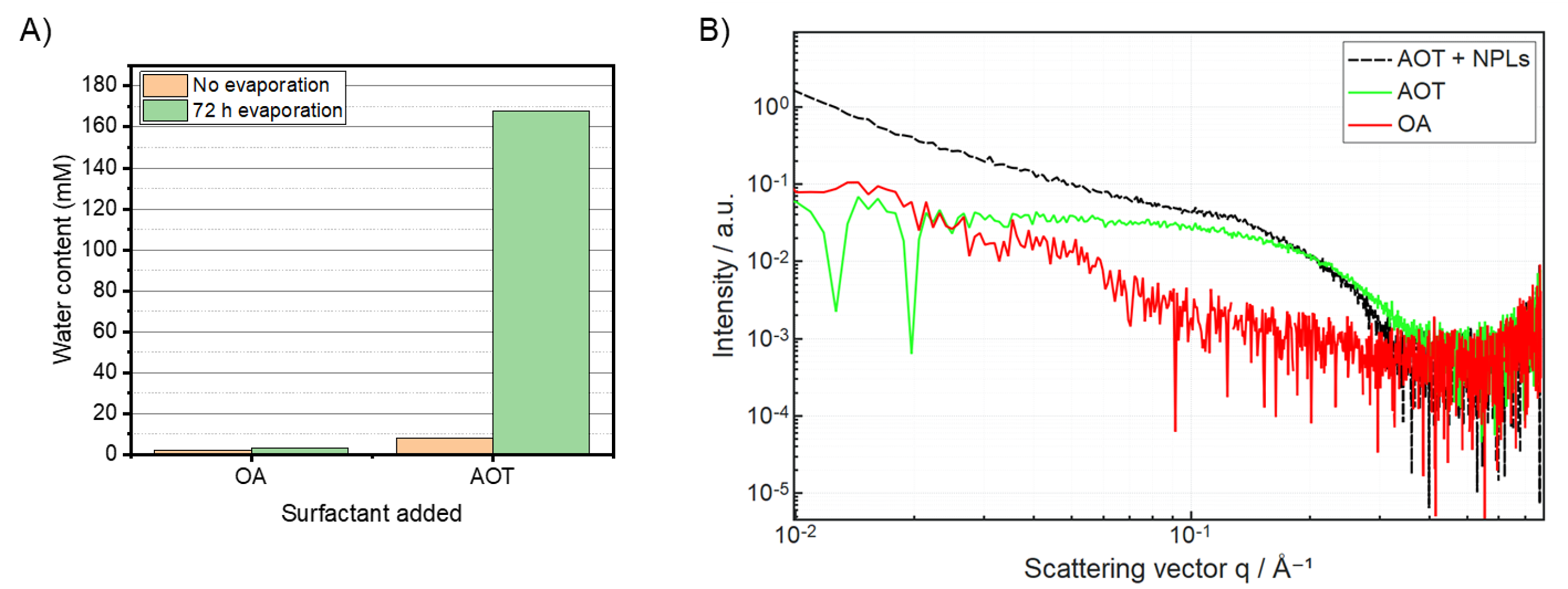}
\caption{A) Water content in a 71 mM surfactant (OA or AOT) in hexane solution before and after 72 h evaporation at 90\% RH. In both cases, more water is measured when AOT is used, and the difference is enhanced after evaporation, implying a stronger hygroscopy. This indicates, along with SAXS data in figure \ref{Depletion_interaction} that AOT forms inverted micelles that are likely more swollen than for their OA counterpart. B) SAXS diagram of 70 mM AOT (green) and OA (red) in hexane. The SAXS diagram corresponding to the self-assembly attempt using AOT is displayed in a dashed black line. The presence of spherical objects is confirmed, and likely related to the presence of AOT reverse micelles.}
\label{Water_uptake_AOT}
\end{figure}

\begin{figure}[htbp]
\centering
\includegraphics[width=1\textwidth,keepaspectratio=true]{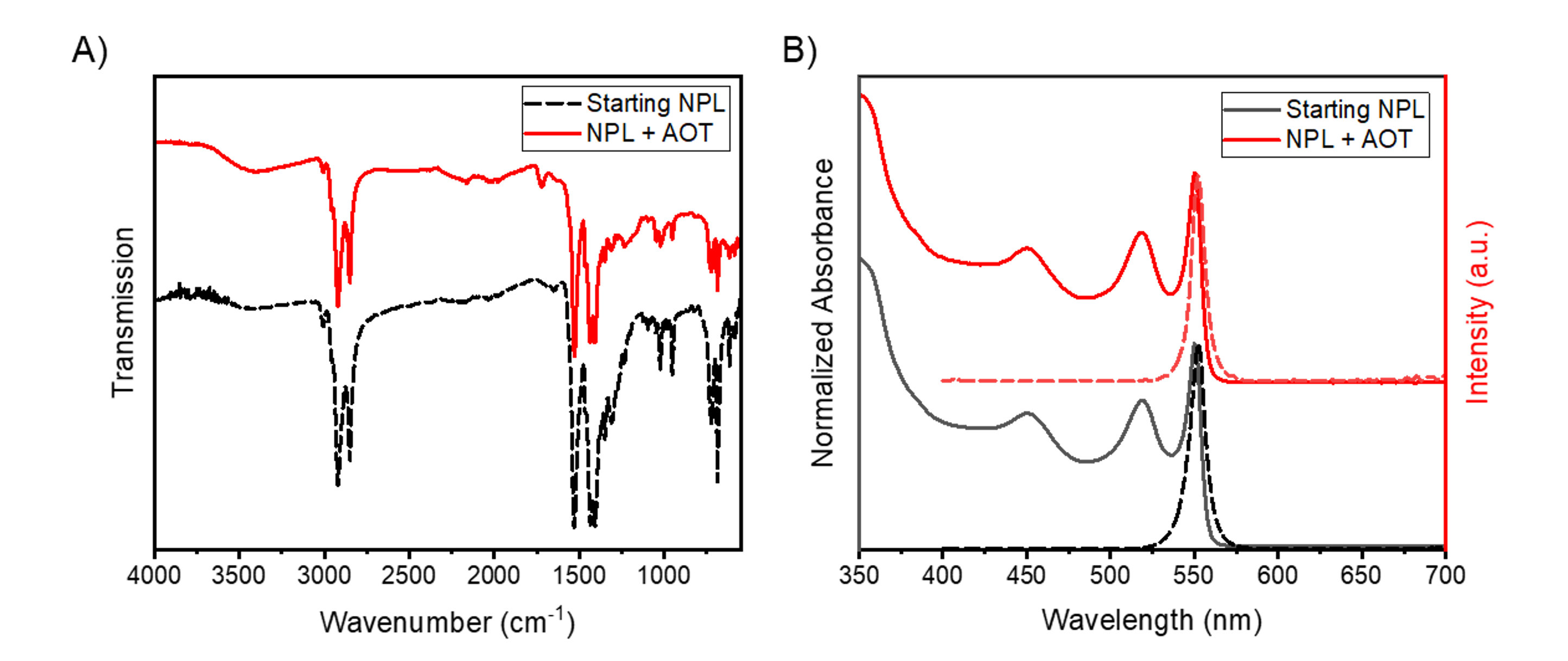}
\caption{A) Corresponding FT-IR spectra on the dried oleate-capped NPL with (red line) or without (dashed black line) 70 mM AOT. B) Corresponding UV-Vis spectra of oleate-capped NPL in hexane suspension with (red line) or without (dashed black line) 70 mM AOT. The lines are superposed in the two graphs for clarity.}
\label{AOT_hexane}
\end{figure}

\begin{figure}[htbp]
\centering
\includegraphics[width=0.9\textwidth,keepaspectratio=true]{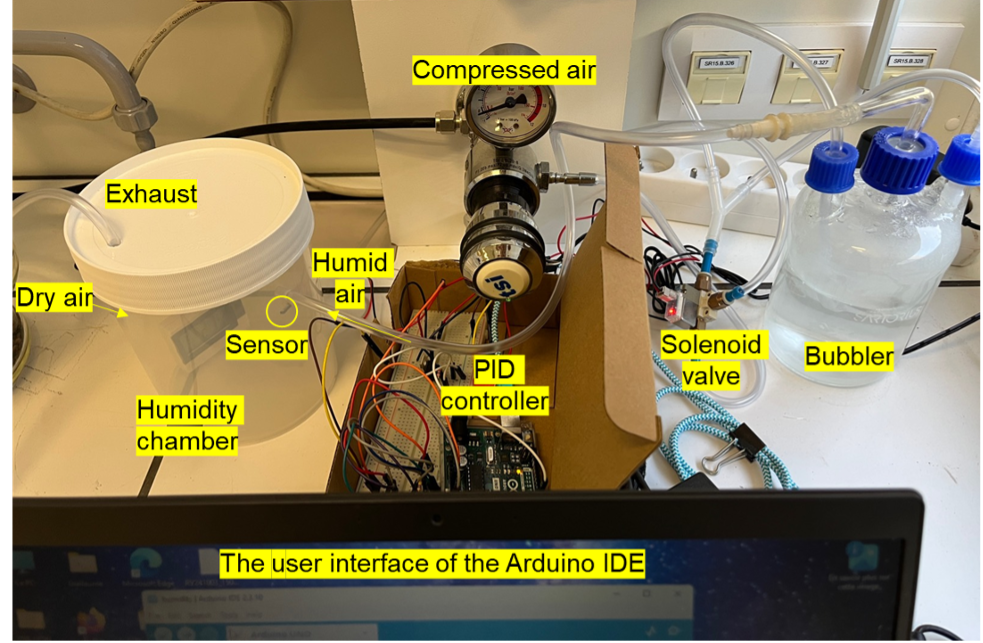}
\caption{A photograph of the humidity regulator set-up featuring a PID controller integrated with the humidity sensor, solenoid valve, and solder-less breadboard. Upon setting the targeted relative humidity via the Arduino IDE, the compressed air flows through the solenoid valve manifold, either bubbling through water or supplying dry air under PID control to achieve the set moisture level.}
\label{fig:RH_regulator_V2}
\end{figure}

\begin{figure}[htbp]
\centering
\includegraphics[width=0.9\textwidth,keepaspectratio=true]{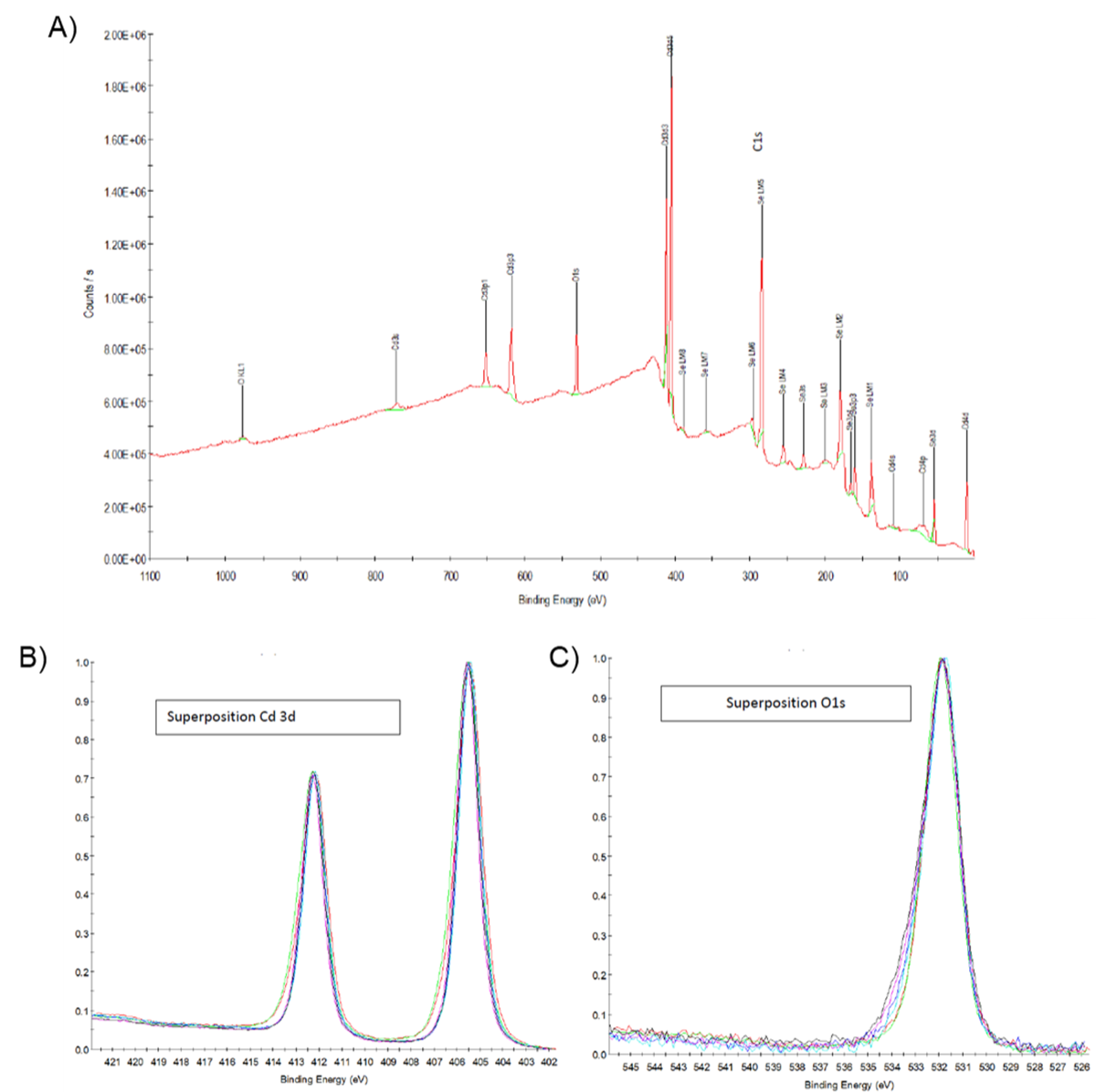}
\caption{A) X-ray photoelectron spectrometry (XPS) of threads samples with close up on B) the Cd 3d region and C) the O 1s region. A superposition with unstacked NPL is made in the two close-ups, but no drastic differences can be noted.}
\label{XPS_SA}
\end{figure}

\end{document}